\documentclass[a4paper,12pt]{article}

\usepackage{amsmath}\usepackage{amssymb}
\usepackage{latexsym}\usepackage{cite}

\newcommand{\be}{\begin{equation}}
\newcommand{\ee}{\end{equation}}

\newcommand{\ka}{\kappa}
\newcommand{\mcA}{{\mathcal A}}
\newcommand{\mcD}{{\mathcal D}}
\newcommand{\mcF}{{\mathcal F}}
\newcommand{\mcL}{{\mathcal L}}
\newcommand{\mcN}{{\mathcal N}}
\newcommand{\mcO}{{\mathcal O}}
\newcommand{\mcV}{{\mathcal V}}
\newcommand{\mcW}{{\mathcal W}}
\newcommand{\half}{\frac{1}{2}}
\def\Dbarhat{\hat{\makebox[0pt][c]{\raisebox{0.5pt}[0pt][0pt]{$\not$}}\mcD}}
\def\Fbarhat{\hat{\makebox[0pt][c]{\raisebox{0.5pt}[0pt][0pt]{$\not$}}F}}

\def\beq{\begin{equation}}
\def\eeq{\end{equation}}
\def\pl{\partial}

\def\al{\alpha}
\def\bt{\beta}

\def\ga{\gamma}
\def\de{\delta}

\def\ka{\kappa}
\def\si{\sigma}
\def\Si{\Sigma}
\def\te{\theta}
\def\La{\Lambda}
\def\lam{\lambda}
\def\Om{\Omega}
\def\om{\omega}
\def\ep{\epsilon}
\def\ze{\zeta}

\def\sq{\sqrt}

\def\l{\left (}
\def\r{\right )}

\def\fr{\frac}
\def\la{\label}
\def\hs{\hspace}
\def\vs{\vspace}

\def\ov{\overline}
\def\tl{\tilde}

\def\un{\underline}

\begin{document}

\begin{titlepage}
\begin{flushright}
HD-THEP-04-24\\
August 18, 2004
\end{flushright}
\vspace{0.6cm}
\begin{center}
{\Large \bf Superfield Approach to 5D Conformal SUGRA\\
 and the Radion$^{\sharp }$} 
\end{center}
\vspace{0.5cm}

\begin{center}
{\large   
Filipe Paccetti Correia$^{\diamond }$,
Michael G. Schmidt$^{\diamond }$,
Zurab Tavartkiladze$^{\diamond }$}

\vspace{0.3cm}

{\em 
Institut f\"ur Theoretische Physik,
Universit\"at Heidelberg\\ 
Philosophenweg 16, 69120 Heidelberg, Germany}
\end{center}
\vspace{0.4cm}
\begin{abstract}

We propose that the radion chiral supermultiplet of five dimensional 
compactified supergravity  is obtained by reduction 
of the graviphoton gauge multiplet to $\mcN=1$ superfields in the off shell
5D superconformal gravity formalism of Fujita, Kugo and Ohashi. 
We present a superfield Lagrangian of Chern-Simons type (similar to 
global SUSY), which reproduces all component couplings of  
gauge fields and the radion. A hypermultiplet superspace action 
is also proposed which correctly accounts for the coupling of matter multiplets
with gauge and radion superfields. 4D supergravity enters by the coupling
to the 4D Weyl multiplet, an even orbifold parity multiplet embedded in the 5D Weyl multiplet. We apply this formalism to a discussion of Fayet-Iliopolous terms, and the gauging of orbifold SUGRA to obtain warped solutions.

\end{abstract} 
\vspace{8cm}
\footnoterule

{\small
\noindent$^\sharp $Preliminary results presented at {\it Seventh European Meeting} 
Planck'$04$, Bad Honnef, Germany, May 24-28, 2004.\\
\noindent$^\diamond $E-mail addresses: F.Paccetti@, M.G.Schmidt@,
Z.Tavartkiladze@ThPhys.Uni-Heidelberg.DE} 

%
%\vspace{4.0cm}
%\begin{flushleft}
%E-mail: 
%
%\end{flushleft}

\end{titlepage}

\section{Introduction}

Five dimensional supergravity (5D SUGRA) today has become a convenient
testground studying the effects of higher dimensions predicted by
string/$M$
theory. Going beyond the original approach to 5D SUGRA 
\cite{cre, gun} one now
preferably considers compactification on an orbifold leading to the
'braneworld' picture \cite{hor,antoni}, that is, one just
compactifies on a folded circle $S^1/Z_2$ with a radius $R$ in the fifth
dimension. Alternatively one can also try to obtain braneworlds dynamically through BPS domain wall constructions \cite{BPS}. The Randall-Sundrum (RS) \cite{ran1} warped solution for such  
settings is quite inspiring, allowing for a fresh view of the hierarchy problem. A $S^1/Z_2$ subspace also appears in Ho\v rava-Witten
theory after compactifying six extra dimensions on a Calabi-Yau manifold \cite{hor, luk}. 
If our world
is located on one of the two branes (4D boundaries), this picture gives new
insights in SUSY breaking and flavor physics if one inspects the 5D
'bulk mediated' interaction with SUSY being broken on a 'hidden' brane
\cite{rat}. It also might lead to successful models for inflation,
relating the inflaton field to a Wilson line in the fifth dimension
\cite{ark00,ginfothers,hof} and a stabilization of the radius 
(associated with a new 'radion' degree of freedom). 

In the commonly assumed orbifold scenario, besides the fields in the 5D bulk, there
are also fields introduced (or localized) on the branes and we need a
consistent treatment of 5D and 4D SUGRA and of other gauge and matter
supermultiplets. Of course, finally only the effective 4D theory obtained
by integrating out the effects of the fifth dimension is of
phenomenological interest. As argued in the simplified case of global SUSY
by Mirabelli and Peskin \cite{mir} one should use the 'off shell' version
of SUSY, i.e. keep the auxiliary fields in order to obtain in a consistent
and elegant way the brane-bulk interaction. We will follow that line
\cite{zucker,kugo00,kugo01,fuji01,kugo01a,kugo02,bergs01,bergs02}. We should however emphasize that also the on shell
formulation \cite{fal, lal, alt} together with some consistency
conditions is quite manageable. 

Off shell 5D SUGRA has been pioneered by Zucker \cite{zucker}. His work has been 
used in most phenomenological discussions up to now 
\cite{ger1, ger2, ghe, rat, buc}.
The advantage of this approach is that tensor calculus can be used, that
the gauge algebra closes, that we can first design symmetry
transformations and then construct
the Lagrangian, and that one can use BRST formulation for
quantization. This approach was improved later on performing a reduction
from 6D superconformal tensor calculus also inducing matter-Yang-Mills
systems \cite{kugo00, kugo01} and using hypermultiplet 
compensators instead
of tensor ones in \cite{zucker}. It still requires a number of field
redefinitions, but then indeed leads to old minimal Poincare supergravity
\cite{ste} on the 4D boundary. The 5D superconformal tensor calculus was
then brought into its final form by Fujita, Kugo and Ohashi in a series of
papers \cite{fuji01, kugo01a, kugo02}  quoted 
as FKO in the following. It contains a
strict conformal gauge fixing and properly discusses the $U(1)_R$ gauging
based on the work of ref.\cite{ber} for the on-shell formulation.

If one wants to consider inter-brane distance fluctuations, one is lead to the
radion-field $R(x)$. Supersymmetrization naturally leads to a radion chiral
superfield \cite{lut} in 4D notation 
$$
T=(R+{\rm i}A_y,~\psi_{y-},~F_T)~,
\la{Tsup}
$$
where $A_y$
and $\psi_{y-}$ are the fifth components of the graviphoton and of the
(projected) gravitino field (appropriately rescaled with $\ka=M^{3/2}_5$). 
Such a superfield can be identified in
Zucker's formulation of 5D SUGRA in the gravitational multiplet
\cite{rat, ger1, ger2} projected to positive 
(zero mode) braneworld parity,
with $F_T$ being a combination of auxiliary fields in this theory. 
In the FKO 5D superconformal formulation \cite{kugo01, kugo01a, kugo02} 
the graviphoton
field is in a separate gauge multiplet. Hence it is clear that we have
to proceed differently in order to identify the radion superfield. 
In section 2 and in the appendices we will shortly discuss 
the FKO
construction and present their action, including powers of the
5D Planck mass, $M_5$, for convenience. In section 3 we will propose and
discuss a superspace Lagrangian for the 5D vector multiplet 
by inspection of the reduction of the 5D vector multiplet to 4D
supermultiplets presented in \cite{kugo02}. In addition we will be
able to identify the \emph{radion multiplet}. Section 4 contains a
discussion of Fayet-Iliopolous terms in 5D SUGRA using the
formalism presented in the preceeding section. Hypermultiplets will be
delt with in section 5. There we present superspace Lagrangians for
hypermultiplets including couplings to vector multiplets of both
positive and negative orbifold parities. We then apply this in section 6
 to gauge the 5D SUGRA in the orbifold. In particular we obtain the RS warped solution.
The way the couplings of the 4D Weyl multiplet with matter arise is 
then briefly discussed in section 7.
We conclude in section 8 with several remarks. The paper contains also four appendices. In appendices A and B, we present the component action of FKO and the superconformal constraints, respectively. Some conventions and superfield expressions are presented in appendix C. Finally, appendix D deals with warped backgrounds.

\section{Off-Shell 5D Supergravity: an Overview}
The construction of off-shell local supersymmetric 5D theories using the framework of conformal supersymmetry \cite{kugo00,kugo01,fuji01,kugo01a,kugo02,bergs01,bergs02} proceeds in the following way: Instead of considering only local supersymmetry and local Poincar\' e transformations, one considers an enlarged set of local transformations, which is obtained by \emph{grading} the algebra of conformal transformations. In this way, in addition to translations (${\bf P}_a$) and Lorentz transformations (${\bf M}_{ab}$), one has dilatations (${\bf D}$) and special conformal transformations (${\bf K}_a$), and besides supersymmetric transformations (${\bf Q}^i$) ($i=1,2$) one has so-called \emph{special} supersymmetric transformations (${\bf S}^i$). There is also an SU(2)$_R$ symmetry (${\bf U}_{ij}$) under the which the fermionic generators transform as doublets, the bosonic ones as singlets. The corresponding gauge fields are:
\be
                      e_{\mu}^{\,a},~\omega_{\mu}^{\,ab},~b_{\mu},~ f_{\mu}^{\,a},~\psi_{\mu}^{\,i},~\phi_{\mu}^{\,i},~V_{\mu}^{\,ij}.
\ee

The number of bosonic degrees of freedom (d.o.f.) exceeds by far the number of fermionic ones. This, and the fact that the symmetries are internal symmetries which are in no connection with the reparametrizations of the manifold, makes it necessary to impose a set of constraints which make the fields $\omega_{\mu}^{\,ab},~\phi_{\mu}^{\,i},~ f_{\mu}^{\,a}$, dependent from the other gauge fields. These unconstrained fields, plus the auxiliary fields needed to close the algebra off-shell, build the so-called \emph{Weyl}-multiplet:
\be
                      (e_{\mu}^{\,a},~b_{\mu},~\psi_{\mu}^{\,i},~V_{\mu}^{\,ij},~v_{ab},~\chi^i,~D),
\ee
where $v_{ab}$ is a real anti-symmetric boson, $\chi^i$ is an SU(2)$_R$ Majorana fermion, and $D$ is a real scalar. This multiplet has $(32+32)$ d.o.f., which is less than the $(48+48)$ needed to build a physically consistent theory. Compensator multiplets must be introduced, which account for the missing degrees of freedom. These are (in the minimal version) a $U(1)$ vector multiplet (${\mathbb V}^0$) and a hypermultiplet (${\mathbb H}^{\alpha}$). While the vector multiplet fixes the superconformal symmetries (${\bf D},~{\bf K}_a,~{\bf S}^i$) down to Poincar\' e supersymmetry, the hypermultiplet fixes the SU(2)$_R$ symmetry.

In addition to the compensator vector multiplet ${\mathbb V}^0$ one may couple $n_V$ further vector multiplets ${\mathbb V}^I$ to SUGRA ($I=1,\dots,n_V$). An off-shell 5D vector multiplet consists of a scalar $M$, an SU(2)$_R$ doublet of fermions $\Omega^i$, a gauge field $W_{\mu}$ and an SU(2)$_R$ triplet of auxiliary scalars $Y^{ij}$:
\be
                       {\mathbb V}^I=(M,~\Omega^i,~W_{\mu},~Y^{ij})^I.   
\ee
All fields transform in the adjoint representation of the gauge group $G$, so that for instance $M=M^I t_I$ where $\{t_I\}$ are the generators of the gauge group.\footnote{Here the $t_I$ are hermitian. The results of FKO are 
obtained with $t^I=-it^I_{\textup{FKO}}$ and 
$[A,B]^I=-i[A,B]^I_{\textup{FKO}}$.} 
The fixing of the ${\bf D},~{\bf S}^i,~{\bf K}_a$ is achieved by 
imposing constraints on the scalars and on the gauginos (see also 
appendix \ref{constraints}):
\be
\mcN (M)=\kappa^{-2},\quad \mcN_I(M)\Omega^{I}=0,
\quad {\hat\mcD}_{a}\mcN (M)=0,
\la{consrev}
\ee
where $\kappa^{-2}\equiv M_5^3$ and the \emph{norm function} $\mcN(M)$ is given by
\be
                      \mcN (M)=\kappa c_{IJK} M^I M^J M^K. 
\ee 
Here $I,J,K=0,\dots,n_V$, and the coefficients $c_{IJK}$ are real and totally symmetric.

The (off-shell) 5D hypermultiplet ${\mathbb H}^{\al }$ consists of two scalars ${\cal A}^{\al }_i$, a Dirac spinor $\ze^{\al }$, and two auxiliary fields ${\cal F}^{\al }_i$:
\beq
{\mathbb H}^{\al }=({\cal A}^{\al }_i,~\ze^{\al }, {\cal F}^{\al }_i)~.
\la{hypmult}
\eeq
Here $i=1, 2$ is the $SU(2)_R$ index. The superscript $\al $
has an even number of values, $\al=1, 2, \cdots , 2r$, and
describes the representation of a subgroup $G'$ of the gauge group
$G$ to which ${\mathbb H}$ couples. $G'$ includes the $U(1)$ gauge groups 
to which we will restrict for simplicity in our work.
Among the considered $U(1)$ gauge groups there is the $U(1)_Z$ 
corresponding to the graviphoton gauge supermultiplet, ${\mathbb V}^0$, which gauges the 
central $Z$-charge. The hypermultiplet gives an infinite dimensional representation of the $U(1)_Z$. 
The index $\al $ is raised and lowered with a $G'$ invariant tensor
$\rho_{\al \bt } $  
($\rho^{\ga \al }\rho_{\ga \bt }=\de^{\al }_{\bt }$).
At least one hypermultiplet is unphysical - 
it is a compensator, needed for gravity to
have canonical form and to fix the $SU(2)_R$ symmetry. 
In order to clarify the notation, 
let us consider the kinetic term for the lowest scalar components
$D_{\mu }{\cal A}^{\bar{\al }}_iD^{\mu }{\cal A}_{\al }^i=
D_{\mu }{\cal A}^{\al }_id_{\al }^{\bt }D^{\mu }{\cal A}_{\bt }^i$, where
$d_{\al }^{\bt }$ is a metric matrix. 
The scalar components satisfy the reality condition
\beq
({\cal A}_{\al i})^*={\cal A}^{\al i}=
\rho^{\al \bt }\ep^{ij}{\cal A}_{\bt j}~,
\la{genrel}
\eeq
and similar for the ${\cal F}$ components. 
In the standard representation we have \cite{wit}
\beq
d={\rm Diag ({\bf 1}_{2p},~-{\bf 1}_{2q})}~,~~~~\rho =\ep \otimes {\bf 1}~, 
\la{strep}
\eeq
where ${\bf 1}_{2p}$ corresponds to the compensators, while ${\bf 1}_{2q}$ to the physical hypermultiplets. For the former (as FKO)
we use the index $\un{\al }$, for the latter  $\tl{\al }$. 
In this way the compensator hypermultiplet will be denoted by ${\mathbb H}^{\un{\al}}$
and the physical one by ${\mathbb H}^{\tl{\al }}$.
With these conventions the kinetic term of 
the scalar components will have the form
$D_{\mu }{\cal A}^{\bar{\al }}_iD^{\mu }{\cal A}_{\al }^i=
-|D_{\mu }{\cal A}_{\un{\al }i}|^2+|D_{\mu }{\cal A}_{\tl{\al }i}|^2$ and 
one sees that the compensators ${\cal A}_{\un{\al }i}$ are unphysical because of their negative 
kinetic terms. As one can read from the Lagrangian, eq.\eqref{L_aux} 
in appendix \ref{sec:actionsFKO}, after integrating out the auxiliary field $D'$, the
coupling $D'({\cal A}^2+2\mcN)$ will impose the
constraint ${\cal A}^2=-2\mcN=-2\ka^{-2}$  on the hypermultiplets. This VEV breaks the $SU(2)_R$ as advertised before.

The field content we just described can be found in 
appendix \ref{sec:actionsFKO}, 
where the fields are classified according to their orbifold 
parities\footnote{Besides 
the multiplets mentioned above, linear and tensor
multiplets can be introduced in 5D SUGRA. These multiplets and their
properties (allowing to embed vector and hypermultiplets into them)
turn out to be very useful and powerful for building invariant actions
\cite{kugo00,kugo01,bergs01,fuji01,kugo01a,kugo02}. However, we do not need to consider these multiplets here.}.
The off-shell component action for 5D SUGRA of FKO can be found in 
the same appendix.

%%%%%%%%%%%%%%%%%%%%%%%%%%%%%%%%%%%%%%%%%%%%%%%%%%%%%%%%%%%%%%%%%%%%%%%%%%%%%%%%%%%%%%%%%%%%%%%%%%%%%%%%%%%%%%%%%% VECTOR MULTIPLETS %%%%%%%%%%%%%%%%%%%%%%%%%%%%%%%%%%%%%%%%%%%%%%%%%%%%%%%%%%%%%%%%%%%%%%%%%%%%%%%%%%%%%%%%%%%%%%%%%%%%%%%%%%%%%%%%%%%%%

\section{Vector Multiplets: ${\mathcal N}=1$ Supermultiplets and Superspace Action}
It is a well known fact that from a 4D point of view 5D ${\mathcal N}=1$ supersymmetry corresponds to ${\mathcal N}=2$ supersymmetry and that multiplets of rigid 5D supersymmetry reduce to pairs of (rigid) 4D supermultiplets. This also means that in the rigid case the components of the 5D supermultiplets can be assembled in pairs of superfields and one can use all the power of superspace to write down 5D supersymmetric actions in a rather straightforward way, as was done in \cite{marc83,ark01,marti01,heb01,dudas04}. On the other hand, a systematic study of the reduction of multiplets of (local) 5D conformal supersymmetry to 4D ones was given in \cite{kugo02}, with the intention to formulate interaction terms on the branes. Here we recall the results for the vector multiplet and identify the \emph{radion multiplet}. Using these $\mcN=1$ multiplets we then write down an 5D action for the Abelian vector multiplets, including the radion multiplet. The hypermultiplet will be handled in section \ref{sec:hyper}.

\subsection{Reduction of the vector multiplet and radion multiplet}
Before we proceed a word on the notation. In this paper we will use two different ways of representing the supermultiplets of $\mcN=1$ supersymmetry. The one is a component notation where the fermions are four-component Majorana spinors. It has the advantage that it is the one used in refs.\cite{kugo02,kak}, where rules are given for the multiplication of multiplets and the construction of actions invariant under 4D superconformal symmetries. The other is the superfield notation with two-component Weyl spinors, which is rather usefull in applications where one doesn't focus on 4D conformal gravity. These two notations are, of course, equivalent and the way one switches between them is explained in appendix \ref{sec:spinors}.
  
The 5D vector multiplet reduces to a 4D gauge multiplet $V^I\equiv(A_{\bar\mu}^I,\lambda^I,D^I)$ plus a chiral multiplet $\Sigma^I\equiv(\phi^I,\chi^I,F^I_{\phi})$. The vector multiplet has Weyl and chiral weights $(w,n)=(0,0)$ and is given by \cite{kugo02}
\be
           V^I=(W_{\bar\mu},2\Omega_+,2Y^{3}-{\hat {\mathcal D}}_5 M)^I,       \la{redV4d}
\ee
where 
\be
          \Omega_+\equiv \Omega_R^1+\Omega_L^2,        
\ee
and the covariant derivative of $M$ is
\be\label{eq:covderiv}
          {\hat {\mathcal D}}_{5} M=(\partial_{5}-b_{5})M-ig[W_{5},M]-2\ka i{\bar \psi}_{5}\Omega.
\ee
(After fixing the dilatation symmetry one has $b_{5}=0$, 
see appendix \ref{constraints}). Note that since we are going to consider Abelian vector multiplets only, the commutator in the last expression vanishes. In the rigid limit ($\ka\to 0$) the gravitino term drops in \eqref{eq:covderiv} and $V^I$ becomes the vector multiplet identified in \cite{mir}.

The chiral multiplet $\Sigma^I=(\phi^I,\chi^I_R,F^I_{\phi})$, with weights $(0,0)$, is given by \cite{kugo02}:
\be
           \begin{array}{ccl} \phi^I & = & \frac{1}{2}(e^5_y M^I-iW_y^I),\\
                              \chi^I & = & 2e^5_y\gamma_5\Omega_{-}^I-2i\ka\psi_{y-}M^I,\\
                              F^I_{\phi} & = & -e^5_y(Y^1+iY^2)^I-i\kappa M^I(V^1_y+iV^2_y)+i\ka{\bar\psi}_{y-}(1+\gamma_5)\Omega_{-}^I, \end{array}         
\la{redSi4d}
\ee
where 
\be
          \Omega_{-}\equiv i(\Omega_R^2+\Omega_L^1).
\ee

A rather interesting object arises if one contracts $\Sigma^I$ with ${\mathcal N}_I(M)$: ${\Sigma}_T\equiv (\mcN_I/3\kappa\mcN)\Sigma^I$. Using the constraints ${\mathcal N}=\kappa^{-2}$, ${\mathcal N}_I\Omega^I=0$ and the fact that ${\mathcal N}_I M^I=3{\mathcal N}$, one gets for its components
\be
           \begin{array}{rcl} {\phi}_T & = & \frac{1}{2}(\kappa^{-1} e^5_y -i{B}_y),\\
                              {\chi}_T & = & -i2\psi_{y-},\\
                              {F}_{\phi_{\begin{tiny}T\end{tiny}}} & = & e^5_y({t}^1+i{t}^2)-i(V^1_y+iV^2_y). \end{array}         
\ee
Here $B_y\equiv (\mcN_I/3\kappa\mcN) W_y^I$ and ${\vec t}\equiv -(\mcN_I/3\kappa\mcN) {\vec Y}^I$. Following literature on the subject, this may be called the \emph{radion supermultiplet} even though only in the $\kappa\to 0$ limit it becomes a supermultiplet (of weights $(-1,0)$). However, this doesn't need to bother us, as the couplings involving its components arise from the superspace action that we will present below without need to introduce a radion superfield separately. To see what happens in the $\ka\to 0$ limit one must consider the norm function. After suitable redefinitions of the scalars $M^I$, the vacuum is given by $M^0=(c_{000})^{-1/3}\ka^{-1}$, $M^{I\neq 0}=0$. One can therefore perturb around this vacuum, which corresponds to an expansion in powers of $\ka M^{I\neq 0}$. It is then not difficult to find out that
\be
         \frac{{\mathcal N}_I}{ 3\kappa{\mathcal N}}=(c_{000})^{1/3}\delta^0_I+\mcO (\ka M^{I\neq 0}).    
\ee
It is thus clear that in the $\ka \to 0$ limit $\Sigma^0$ is the radion superfield. On the other hand, for $I\neq 0$, $\Sigma^I$ reduces in the $\ka \to 0$ limit to the chiral supermultiplet identified in \cite{mir} up to a pre factor $e^5_y$.

There is still another multiplet $\mcV_5^I$ that can be obtained out of the components of the 5D vector multiplet. This is a general type multiplet
%\footnote{A general type superfield is of the form $\Phi=C+\theta\zeta+{\bar\theta}{\bar\zeta}+\theta^2 H+{\bar\theta}^2 K+{\bar\theta}\sigma^{a}\theta B_a+{\bar\theta}^2\theta\lambda+\theta^2{\bar\theta}{\bar\lambda}+{\bar\theta}^2\theta^2 D$.} 
with $(1,0)$ weights and is given by \cite{kugo02}
\be\label{eq:V_5def}
          \mcV_5^I=(~M,-2i\gamma_5\Omega_-,~2Y^1,~2Y^2,~{\hat F}_{a5}+2\ka v_{a5}M,~\lambda^{\mcV_5},~D^{\mcV_5})^I,
\ee
where
\be
            \lambda^{\mcV_5}=-2{\hat \mcD}_5\Omega_++2\ka i\gamma^a v_{a5}\Omega_- -\frac{i}{4}\ka\gamma_5\chi_+M,
\ee
\be
            D^{\mcV_5}={\hat \mcD}_5({\hat \mcD}_5 M-2Y^3)-\frac{1}{4}\ka^2 DM+\ka v^a_5(2{\hat F}_{a5}+\ka v_{a5}M)+\frac{1}{2}\ka{\bar \chi}_+\Omega_-.
\ee
Note the appearance of the auxiliary fields $v_{ab},\chi,D$ of the Weyl multiplet. In the rigid limit all these fields drop out and $\mcV_5$ can be written as a simple combination of the vector and the chiral multiplets as
\be
                \mcV_5^I|_{\ka=0}=(e^5_y)^{-1}(\Sigma+\Sigma^+)^I-\partial_5 V^I.    
\ee
On the other hand, in the general (local) case to obtain $\mcV_5$ out of $\Sigma$ and $V$ one must lift up $e^5_y$ to a full multiplet ${\mathbb W}_y$ with $(-1,0)$ weights. In ref.\cite{kugo02} such a multiplet was identified, consisting of fields from the 5D Weyl multiplet which don't participate in the 4D Weyl multiplet:
\be
             {\mathbb W}_y=(~e^5_y,-2\ka\psi_{y-},-2\ka V_y^2,~2\ka V_y^1,-2\ka v_{ay},~\lambda^{{\mathbb W}_y},~D^{{\mathbb W}_y}),
\ee  
with
\be
             \lambda^{{\mathbb W}_y}=\tfrac{i}{4}\ka e^5_y\gamma_5\chi_+ +2\phi_{y+}+2\ka^2\gamma_5\gamma^bv_{b5}\psi_{y-},
\ee
\be
             D^{{\mathbb W}_y}=\ka^2e^5_y\left[\tfrac{1}{4}D-(v_{a5})^2\right]-2f_y^5+\tfrac{i}{4}\ka^2{\bar\chi}_+ \gamma_5\psi_{y-},  
\ee
where $\phi_{y+}$ and $f_y^5$ are combinations of the gauge fields of the Weyl multiplet (as pointed out in section 2). We can use this now to write $\mcV_5$ in terms of the superfields $V$ and $\Sigma$ also in the local case
\be\label{eq:V_5defsuper}
               \mcV_5= \frac{\Sigma+\Sigma^+ -\partial_y V}{{\mathbb W}_y} +\cdots   
\ee
This expression misses contributions from some of those fields that belong to the 5D Weyl multiplet but have \emph{negative} parity under orbifolding (see table 1). This means that they belong neither to the 4D Weyl multiplet nor to the radion multiplet. The ratio appearing in eq.\eqref{eq:V_5defsuper} can be calculated with the usual superspace rules, or alternatively one may use the formulas for products of general multiplets of 4D conformal SUGRA as given in \cite{kak,kugo02}. The later differ from the former in that all 4D derivatives become covariant in respect to 4D conformal supergravity. We should emphasize that $\mcV_5$ is invariant under the abelian gauge transformations and can be coupled directly to the orbifold fix-points since it also transforms trivially under the \emph{odd} superconformal symmetries at the boundaries. This is in strong contrast to the behaviour of $\Sigma$ and ${\mathbb W}_y$ which at the branes transform in a non-trivial way under the odd superconformal symmetries and, in the case of $\Sigma$, also under the gauge symmetries.

Finally let us mention the effects of orbifolding the 5th dimension. The requirement of invariance of the action under orbifold projections implies that $W_{\bar\mu}$ and $W_y$ must have opposite orbifold parities, i.e. 
\be
       \Pi(W_{\bar\mu})=-\Pi(W_y).
\ee
This in turn means that 
\be
          \Pi(V^I)=-\Pi(\Sigma^I)=-\Pi(\mcV_5^I).
\ee
The two possible choices, $\Pi(V^I)=+1$ and $\Pi(V^I)=-1$, give two essentially different physical pictures at low energies, in particular the second choice allows the breaking of an $U(1)$ gauge symmetry by orbifolding (see section \ref{sec:hyper}). See table 1 for the detailed assignment of orbifold parities to the fields of the vector multiplet.

%%%%%%%%%%%%%%%%%%%%%%%%%%%%%%%%%%%%%%%%%%%%%%%%%%%%%%%%%%%%%%%%%%%%%%%%%%%%%%%%%%%%%%%%%%%%%%%%%%%%%%%%%%%%%%%%%%%%%%%%%%%%%%%%%%%%%%%%%%%%%%%%%%%%%%%%%%%%%%%%%%%%%%%%%%%%%%%%%%%%%%%%%%%%   ACTION FOR SUPERFIELDS   %%%%%%%%%%%%%%%%%%%%%%%%%%%%%%%%%%%%%%%%%%%%%%%%%%%%%%%%%%%%%%%%%%%%%%%%%%%%%%%%%%%%%%%%%%%%%%%%%%%%

\subsection{Superspace action}

For the discussion of the effective 4D theory and model building a formulation in terms of 4D ${\mathcal N}=1$ superfields is very usefull. In this section we give such a formulation for the \emph{Abelian} vector part of the Lagrangian, including the radion multiplet.

As pointed out before, a 5D abelian vector multiplet reduces in 4D to a vector plus a chiral multiplet. The corresponding superfields, which we denote by $V^I$ and $\Sigma^I$ ($I=0,\dots,n$), transform under the (abelian) gauge transformation as
\be
                \delta V^I=\Lambda^I+\Lambda^{I+},\qquad \delta \Sigma^I =\partial_y\Lambda^I.
\ee
In the rigid limit, out of these two superfields two independent super gauge invariant superfields can be constructed\cite{ark01,heb01,dudas04}
\be
                 {\mathcal W}_{\alpha}^I=-\frac{1}{4}{\bar D}^2D_{\alpha}V^I,\qquad {\mathcal V}_5^I=(\Sigma^I+\Sigma^{I+})-\partial_5 V^I.
\ee
The local (superconformal) version of $\mcV_5$ has already been presented in the previous subsection (eq.\eqref{eq:V_5def}), we are thus left with ${\mathcal W}_{\alpha}$, which is a chiral multiplet of weights $(3/2,3/2)$:
\be
          {\mathcal W}_{\alpha}=(-i2\Omega_{+\alpha},-i({\Fbarhat})_{\alpha}^{\beta}+\delta_{\alpha}^{\beta}(2Y^3-{\hat\mcD}_5 M),~2({\Dbarhat}\Omega_+)_{\alpha}).
\ee

These invariant superfields can now be used to construct the 5D Lagrangian we are searching for. Let us for this purpose introduce the prepotential\footnote{We will keep using $\mcN$ just as a function of the scalars $M^I$, while $P$ will be a function of the superfields $\Sigma^I$ and $\mcV^I_5$.}
\be
                   P(M)\equiv -\frac{1}{2}{\mathcal N}(M), 
\ee
where $\mcN (M)=\kappa c_{IJK}M^I M^J M^K$ is the norm function introduced above, the real coefficients $c_{IJK}$ being totally symmetric. Note that the prepotential is a cubic polynomial, in agreement with the requirement \cite{seiberg} that it be a gauge invariant \emph{at most} cubic polynomial. We have now all the ingredients we need to write down the Lagrangian in terms of superfields, which turns out to be
%\footnote{Note that despite having the same form as eq.(43) of \cite{dudas04} our expression differs in that this reference's prepotential is $P(M)\sim M^2+M^3$ (see below).}
\be\begin{split}\label{eq:vecsuperlagr}
          e_{(4)}^{-1}{\mathcal L}_5= & \frac{1}{4}\int d^2\theta\left(P_{IJ}(2\Sigma){\mathcal W}^{I\alpha}{\mathcal W}_{\alpha}^J-\frac{1}{6}P_{IJK}{\bar D}^2 (V^I D^{\alpha}\partial_y V^J-D^{\alpha}V^I\partial_y V^J){\mathcal W}_{\alpha}^K\right)\\
                  &\qquad {}+\textup{h.c.}+\int d^4 \theta\, {\mathbb W}_y \,2P({\mathcal V}_5).
\end{split}\ee
This Lagrangian agrees with the 5D SUGRA Lagrangian of FKO \cite{kugo01a}
(eq. (\ref{lagFKOrev}) and following) upon the use of the 
constraints (\ref{consrev}).
A number of remarks is now in order. The first one concerns the Weyl weights: With $w(d^n\theta)=n/2$, one sees that the right hand side of the above expression has Weyl weight four, which indeed compensates for the transformation properties of $e_{(4)}\equiv\textup{det}\,{e_{(4)}}_{\mu}^a$ under dilatations since $w(e_{(4)})=-4$. 
%Related to this is the issue of the extension of the 4D density $e_{(4)}$ to a superfield (of weights $(-4,0)$), usually denoted by ${\mathcal E}$, which we will leave for section 6. Let us just note that ${\mathcal E}$ is build out of the 4D Weyl multiplet, which is obtained by reduction of the 5D one, and enters the action as a pre factor that multiplies all terms, for example in
%\be
%                  \int d^4 \theta\, {\mathcal E}{\mathbb W}_y \,2P({\mathcal V}_5).    
%\ee
Another comment concerns the fact that ${\mathbb W}_y$ and $\Sigma$ transform in a non-trivial way \cite{kugo02} under the \emph{odd} 5D superconformal transformations, unlike, for instance, what happens with ${\mathcal V}_5$. To compensate for this non-trivial behaviour one may need to add terms to ${\mathbb W}_y$, which include derivatives of the corresponding gauge fields, build out of those components of the 5D Weyl multiplet which are odd under orbifold parity. In the same way, one expects the derivative $\partial_y$ acting on $V$ to be promoted to a superoperator including odd elements of the 5D Weyl multiplet, to ensure 5D superconformal invariance.

%This Lagrangian should agree with FKO upon the use of the constraints ${\mathcal N}(M)=\kappa^{-2}$ and ${\mathcal N}_I(M)\Omega^I=0$. Note that even though we consider only abelian vector multiplets, it is straightforward to extend this to the non-abelian case too. Here we assume that the background is flat. Additional interactions involving the fields of the vector multiplets arise from couplings to the Weyl multiplet. In particular self-couplings appear after integrating out the auxiliary anti-symmetric $v_{ab}$, see eq.\eqref{eq:va_ab_sol}.

Let us see now what emerges in the $\ka\to 0$ limit. For this purpose we consider the simple case (see appendix B) that 
\be\label{eq:normf1}
             \ka^{-1}\mcN=\alpha {M^{0}}^3-\beta\alpha^{\frac{1}{3}} M^0 {M^1}^2-\gamma{M^1}^3,
\ee
with $\alpha=(2/3)^{\frac{3}{2}}$ and $\beta=1$. One vacuum is given by $M^0=\alpha^{-1/3}\ka^{-1}$ and $M^1=0$. Expanding around this configuration one gets
\be\label{eq:P_example1}
               P(\varphi)\simeq -\frac{1}{2\ka^2}+\frac{1}{2}\varphi^2+\frac{\gamma}{2}\ka\varphi^3+\cdots,
\ee
where $\varphi={\mathcal V}_5^1,~2\Sigma $, and the dots indicate additional terms involving higher powers of $\kappa M^1$, and couplings to fields of the gravitational sector. One sees that the prepotential becomes the cubic function discussed by Seiberg in \cite{seiberg} and  used in \cite{dudas04}. In this way, in the $\ka\to 0$ limit we obtain the results of \cite{ark01,dudas04}, the supersymmetric Chern-Simons term included. Note, however, that if we put the theory on the orbifold $S^1/Z_2$, the coupling $\gamma$ must become odd, i.e. $\gamma\sim\epsilon (y)$. This has the consequence that under a gauge transformation the above Lagrangian is not invariant, having a non-vanishing transformation on the branes:  
\be
                \delta (e_{(4)}^{-1}{\mathcal L}_5)\sim\kappa (\partial_y \gamma)\int d^2\theta \Lambda^1{\mathcal W}^{1\alpha}{\mathcal W}_{\alpha}^1+\textup{h.c.}=2\kappa|\gamma|\,[\delta(y)-\delta(y-\pi R)]\int d^2\theta \Lambda^1{\mathcal W}^{1\alpha}{\mathcal W}_{\alpha}^1+\textup{h.c.}
\ee   
In fact, it is well known that this can be used to cancel anomalies arising at the fix-point branes \cite{ark01}.

Let us now see explicitely how (for $\ka\to 0$) the \emph{radion sector} couples to the gauge sector. For this recall our remark in section 3.1 that in the $\ka\to 0$ limit $\Sigma^0$ is (proportional to) the so-called radion chiral superfield, $\Sigma_T$. If we introduce $T\equiv 2\ka {\Sigma}_T$ we have in this limit
\be
              \Sigma^0=\frac{\alpha^{-\frac{1}{3}}}{2\ka}T.
\ee 
On the other hand it is not dificult to recognize that there is some overlap between $\frac{1}{2}(T+T^+)$ and ${\mathbb W}_y$, the two superfields becoming identical if the auxiliary fields $v_{ay}$, $\lambda^{{\mathbb W}_y}$ and $D^{{\mathbb W}_y}$ as well as ${\vec Y}^0$ and ${\hat F}^0_{a5}$ (the graviphoton's field-strength) are set to zero\footnote{The fields $\lambda^{{\mathbb W}_y}$ and $D^{{\mathbb W}_y}$ are Lagrange multipliers that can safely be put to zero if one imposes \emph{by hand} the constraints that they imply.}. We have thus
\be
              {\mathbb W}_y=\frac{T+T^+}{2}+\cdots
\ee
In this case one also has $(\Sigma^0+\Sigma^{0+}-\partial_y
V)=\alpha^{-\frac{1}{3}}(2\ka)^{-1}(T+T^+)$ and therefore
$\mcV_5^0=\alpha^{-\frac{1}{3}}\ka^{-1}$. If we use all this we obtain (see
eq.\eqref{eq:P_example1}) 
\be              2{\mathbb W}_y
P(\mcV_5)\simeq-\kappa^{-2}\frac{T+T^+}{2}+2\frac{(\Sigma^1+\Sigma^{1+}-\partial_y V^1)^2}{T+T^+}+4\kappa\gamma\frac{(\Sigma^1+\Sigma^{1+}-\partial_y V^1)^3}{(T+T^+)^2}.   \ee %where we put the dots to remind the reader that we dropped potentially relevant terms.  
Note that the first term in the r.h.s. was given in
\cite{lut}. The second term has the same form as the one first presented in
ref.\cite{marti01} to couple the vector multiplet with the radion multiplet. 

%However this must be enjoyed with some caution. The point is that $(\Sigma^1+\Sigma^{1+}-\partial_y V^1)$ also includes in its components elements of the radion multiplet, like $\psi_{y-}$ and $V_y^{1,2}$, which give rise to additional couplings between the vector multiplet and the radion multiplet. These couplings (absent in \cite{marti01}) can be \emph{shifted away} by a redefinition of the 4D gauginos and other fields in $\Sigma$ and $V$ which makes $(\Sigma^1+\Sigma^{1+}-\partial_y V^1)$ independent of any gravitational fields (apart from the \emph{radion} $e^5_y$, which may be factored out). Such redefinitions clearly break the SU(2)$_R$ symmetry and should therefore be used only when we don't care about the 5D structure. 

Another coupling between the vector multiplet and the radion multiplet arises from the F-term coupling,
\be
              \frac{1}{4}\int d^2\theta P_{IJ}(2\Sigma){\mathcal W}^{I\alpha}{\mathcal W}_{\alpha}^J=\frac{1}{4}\int d^2\theta\,T\,{\mathcal W}^{1\alpha}{\mathcal W}_{\alpha}^1+\cdots,
\ee
which is a Chern-Simons like term and clearly has the same form as the one in \cite{marti01}.

%%%%%%%%%%%%%%%%%%%%%%%%%%%%%%%%%%%%%%%%%%%%%%%%%%%%%%%%%%%%%%%%%%%%%%%%%%%%%%%%%%%%%%%%%%%%%%%%%%%%%%%%%%%%%%%%%%%%%%%%%%%FI-terms%%%%%%%%%%%%%%%%%%%%%%%%%%%%%%%%%%%%%%%%%%%%%%%%%%%%%%%%%%%%%%%%%%%%%%%%%%%%%%%%%%%%%%%%%%%%%%%%%%%%%%%%%%%%%%%%%%%%%%%
\section{Fayet-Iliopolous Terms in 5D SUGRA}

Now we turn to the discussion of FI terms. In 4D they are allowed (for Abelian gauge groups) and cause either SUSY or gauge symmetry breaking. In 5D orbifolds, the situation is different \cite{ghilen01,barbie02,groot02,martiFI} due to the existence of the chiral superfield $\Sigma=M+iA_5+\cdots$ which accompanies the vector superfield $V$. Indeed, the derivative $\partial_5\Sigma$ can cancel the FI terms localized at the fixed point boundaries, in which case SUSY remains unbroken and $M$ gets a step function-like VEV. To extend this analysis to 5D SUGRA we will use the superspace actions described in the previous section. For simplicity we will assume that the background metric has no dependence on the 5th coordinate $y$. To include warped backgrounds we can use the same approach as in sec.\ref{sec:gaugesugra} and appendix \ref{sec:warped}.

As we have just pointed out, FI terms in 5D orbifolds are equivalent to VEV's of odd scalars $M(y)\sim \xi \epsilon(y)$. To obtain such a vacuum let us consider an $S^1/Z_2$ orbifold and the following norm function \cite{barbie02}:
\be
                \ka^{-1}\mcN =(M^0)^3-M^0(M^1)^2+\half \xi\epsilon(y)(M^0)^2 M^1,
\ee
where $M^1(y)$ is odd and $\epsilon(y)$ is the periodic step-function ($\epsilon(\pi R> y>0)=1$). To see how this norm function leads to FI terms consider the following terms arising from the Lagrangian (eq.\eqref{eq:vecsuperlagr}),
\be\label{eq:vecDterms}
        e^{-1}\mcL_5\supset -\frac{1}{4}\mcN_{IJ}(M)D^I D^J+\half\mcN_I(M)\partial_5 D^I.
\ee
After a partial integration we get
\be
         e^{-1}\mcL_5\supset -\frac{1}{4}\mcN_{IJ}(M)D^I D^J-\half\mcN_{IJ}(M) D^I\partial_5 M^J-\half\xi_I(M)[\delta(y)-\delta(y-\pi R)]D^I,    
\ee
where we introduced
\be
                \xi_I(M)\equiv \xi\partial_I [(M^0)^2 M^1]=\xi[\delta^0_I 2M^0 M^1+\delta^1_I(M^0)^2].
\ee
It is now evident that the choice of the norm function leads to FI terms localized at the orbifold fixed points. As in the case of global supersymmetry, the odd scalar $M^1$ gets a VEV $\sim \xi\epsilon(y)$. The best way to show this is by making the shift \cite{barbie02}
\be
                   M^1\to {\bar M}^1+\frac{\xi}{4}\epsilon(y)M^0,
\ee
in which case the norm function becomes
\be
          \ka^{-1}\mcN\to \ka^{-1}{\bar \mcN}=\left( 1+\frac{\xi^2}{16}\right)(M^0)^3-M^0({\bar M}^1)^2,
\ee
and thus is of the same form as eq.\eqref{eq:normf1} with $\gamma=0$. The corresponding \emph{vacuum} is $M^0=\ka^{-1}( 1+\xi^2/16)^{-1/3}$, ${\bar M}^1=0$, which leads to a non-vanishing VEV of the odd scalar $M^1$
\be
                        \langle M^1(y)\rangle=\frac{\xi}{4( 1+\xi^2/16)^{1/3}}\ka^{-1}\epsilon(y).
\ee

This analysis may be extended to the level of superfields through the shifting of the full superfields \footnote{Note that the consistence of both these redefinitions implies that one must have $\epsilon (y)\partial_y V^0=\partial_y (\epsilon (y)V^0)$, which indeed follows from $\delta(y-y_{fp})V_0=0$ for odd $V^0$.}
\be
                        \Sigma^1={\bar\Sigma}^1+\frac{\xi}{4}\epsilon(y)\Sigma^0,\quad \mcV_5^1={\bar \mcV}_5^1+\frac{\xi}{4}\epsilon(y)\mcV_5^0.
\ee
In terms of the new superfields, the superspace action is then obtained by using $P=-\frac{1}{2}{\bar\mcN}$, which means that all $\epsilon (y)$ and $\partial_y\epsilon (y)$ are shifted away from the vector sector. On the other hand, if some hypermultiplet transforms under the $U(1)_1$ gauged by $W_{\mu}^{I=1}$, it will be affected by the shift getting an odd mass 
\be
         m(y)\sim qg_1^2(\xi/4)\epsilon(y)\langle M^0\rangle=qg_1^2\langle M^1(y)\rangle
\ee
and odd couplings to the graviphoton. 
This step-like profile of the mass can then lead to the 
spontaneous localization of
the even hypermultiplet at one of the branes \cite{groot02, martiFI}.
In the $\ka\to 0$ limit $\Sigma^0$ becomes the radion superfield and we are left with the coupling of the radion superfield to the charged chiral superfields.

Note that recently a detailed analysis of FI-terms in 5D orbifold SUGRA was presented in ref.\cite{abe04}, where the 4-form mechanism of ref.\cite{ber} in its off-shell version \cite{kugo01a} was used. Even though we use a different approach one shouldn't expect any differences in the results.

%%%%%%%%%%%%%%%%%%%%%%%%%%%%%%%%%%%%%%%%%%%%%%%%%%%%%%%%%%%%%%%%%%%%%%%%%%%%%%%%%%%%%%%%%%%%%%%%%%%%%%%%%%%%% HYPERMULTIPLET %%%%%%%%%%%%%%%%%%%%%%%%%%%%%%%%%%%%%%%%%%%%%%%%%%%%%%%%%%%%%%%%%%%%%%%%%%%%%%%%%%%%%%%%%%%%%%%%%%%%%%%%%%%%%%%%

\section{Hypermultiplet Superspace Action}\label{sec:hyper}
As is well known, it is convenient to split the 5D hypermultiplets ${\mathbb H}^{\al }=(\mcA^{\alpha}_i,~\ze^{\al }, {\cal F}^{\al }_i)$ into $r$ pairs
$({\mathbb H}^{2\hat{\al }-1},~{\mathbb H}^{2\hat{\al }})$, where $\hat{\al }=1, 2, \cdots , r$
indicates the number of introduced 5D 
hypermultiplets. The reason for doing this is the reality condition, eq.\eqref{genrel}, which now reads
\beq
({\cal A}_2^{2\hat{\al }})^*={\cal A}_1^{2\hat{\al }-1}~,~~~~
({\cal A}_1^{2\hat{\al }})^*=-{\cal A}_2^{2\hat{\al }-1}~,
\la{relcond}
\eeq
(and similarly for ${\cal F}$ components) and clearly relates ${\mathbb H}^{2\hat{\al }-1}$ with ${\mathbb H}^{2\hat{\al }}$. Therefore, for each 
$\hat{\al }$ only 
four real scalar components are independent\footnote{For a more detailed 
discussion about hypermultiplets see \cite{kugo01, kugo01a}.}.
For a given $\hat{\al}$, the 5D hypermultiplet decomposes into a 
pair of $\mcN=1$ 4D chiral superfields
with opposite orbifold parities \cite{kugo02}:
$$
H=\l {\cal A}^{2\hat{\al }}_2,~-2{\rm i}\ze^{2\hat{\al }}_R,~
({\rm i}M_*{\cal A}+{\hat{\mcD}}_5{\cal A})^{2\hat{\al }}_1\r, 
$$
\beq
H^c=\l {\cal A}^{2\hat{\al }-1}_2=-({\cal A}^{2\hat{\al }}_1)^*,~
-2{\rm i}\ze^{2\hat{\al }-1}_R,~
({\rm i}M_*{\cal A}+{\hat {\mcD}}_5{\cal A})^{2\hat{\al }-1}_1\r, 
\la{hyperdec}
\eeq
with 
$$
M_*{\cal A}^{\al }_i=igM^I(t_I)^{\al }_{\bt }{\cal A}^{\bt }_i+
{\cal F}^{\al }_i~,
$$
\beq
\hat{\mcD}_{5 }{\cal A}^{\al }_i=
\pl_{5 }{\cal A}^{\al }_i-i
gW_{5 \bt }^{\al }{\cal A}^{\bt }_i-
W_{5 }^0\fr{1}{\al }{\cal F}^{\al }_i-
\ka V_{5 ij}{\cal A}^{\al j}-
2\ka {\rm i}\bar \psi_{5 i}\ze^{\al }~,
\la{covders}
\eeq 
where $(t_I)^{\al }_{\bt }$ is the generator of gauge group
$G_I$. Eq.\eqref{hyperdec} should be compared with the usual 
decomposition in global SUSY \cite{mir,heb01}, which we reproduce 
here in the notation of \cite{heb01}
$$
H_{gl}=H^1+\sq{2}\te \psi_L+\te^2(F_1+D_5H^2-\Si H^2)~,
$$
$$
H^{c}_{gl}=
H_2^{\dagger }+\sq{2}\te \psi_R+\te^2(-F^{\dagger 2}-
D_5H_1^{\dagger }-H_1^{\dagger }\Si )  ~,
$$
when we take the limit $\ka \to 0$.
In the following we will use the notation
\beq
{\bf H}\equiv ({\bf H}_1,~{\bf H}_2)=(H, ~H^c)~,
\la{hyper}
\eeq
keeping in mind that if we have more than one 5D hypermultiplet, i.e. for $r>1$, the index $\hat{\al}$ should be present, ${\bf H}^{\hat{\al}}\equiv ({\bf H}_1,~{\bf H}_2)^{\hat{\al}}=(H, ~H^c)^{\hat{\al}}$, but for legibility remains only implicit.

As we have seen in section 3.1, a 5D vector multiplet reduces to a 4D 
vector superfield $V^I$ (eq.\eqref{redV4d})
and a chiral superfield $\Si^I$ (eq.\eqref{redSi4d}). 
We want to build a superspace action for the hypermultiplet 
$({\bf H}_1,~{\bf H}_2)$
interacting with a 5D gauge multiplet $(V, ~\Si)$. 
For this purpose we introduce
\beq 
{\bf V}^{ab}=gV\vec q \cdot \vec \si^{ab}~,~~~~~
{\bf \Si}^{ab}=g\Si\vec q \cdot \vec \si^{ab}~,~~~
{\rm with}~~|\vec q|=1~.
\la{gaugeR}
\eeq
(Like in QED, the coupling constants only appear with matter and are written 
explicitely. In the gauge kinetic part, there are no gauge couplings.)
This notation turns out to be convenient for constructing the action
invariant under different orbifold parity prescriptions for the vector 
superfields. In the component off-shell formulation this notation
has been already
used in order to gauge the $U(1)_R$ symmetry 
\cite{kugo01a}.
Here we use it for a general Abelian $U(1)$ gauge symmetry.
%
%\footnote{Besides
%this, there is an auxiliary gauge field $V_{\mu }^{ij}$, which comes from
%5D Weyl multiplet and couples with $SU(2)_R$ doublets through covariant
%derivatives. Since it is the part of 5D gravity, we will ignore it here 
%believing that superfield formulation including 4D SUGRA should also 
%be possible.}.
%

Now we are ready to write a superspace Lagrangian for hypermultiplets.
In the case of one $r$-hypermultiplet and one gauge field it has the form
\beq
e_{(4)}^{-1}{\cal L}({\bf H})=\int d^4\te\, {\mathbb W}_y 2{\bf H}^{\dagger }_a
(e^{-{\bf V}})^{ab}{\bf H}_b-
\int d^2\te ({\bf H}\ep )_a\l \hat{\pl_y}-
{\bf \Si } \r^{ab}{\bf H}_b
+{\rm h.c.} 
\la{hypAct}
\eeq 
where the superoperator $\hat{\pl }_y$ is obtained by promoting ${\pl }_y$
to an operator containing odd (under orbifold parity) elements of the 5D
Weyl multiplet, namely
$\hat{\pl }_y={\pl }_y+\La^{\al }D_{\al }+\La^{\ov{\mu }}{\pl }_{\ov{\mu
}}$.
The superfields $\La^{\al }$, $\La^{\ov{\mu }}$ are such that
$\hat{\pl }_y{\bf H}_a$ is a chiral superfield. For this $\La^{\al }$
must be a chiral superfield, $\La^{\al }=\ov{D}^2L^{\al }$,
with a spinor index ($L^{\al }$ is a general complex superfield).
The superfield $\La^{\ov{\mu }}$ is related to $L^{\al }$:
$\La_{\al \dot{\al }}=8i\ov{D}_{\dot{\al }}L_{\al }+\Om_{\al \dot{\al }}$,
where $\Om_{\al \dot{\al }}$ is also a chiral superfield. With these conditions
it is straightforward to check out that the $\hat{\pl }_y{\bf H}_a$ is
chiral. This type of construction is important for obtaining the correct interactions
of matter with the 5D Weyl multiplet. The lowest component of $\La^{\al }$
is $\La^{\al}|_{\te =0}=\ka (\psi_{y+})^{\al}_L=\ka(\chi^2)^{\al }$
(see appendix \ref{sec:spinors} for conventions). This term is important for
the cancellation of ${\cal F}\ov{\psi }\zeta $ type terms, in order to recover
the FKO Lagrangian. Higher components of $\La^{\al }$ should be obtained
by SUSY (and superconformal) transformations.

%where $\hat{\pl_y}=\pl_y+\dots$, the dots staying for a (unknown) superfield containing elements of the 5D Weyl multiplet odd under the orbifold parity (see discussion in section \ref{sec:Weylmultiplets}). 
%=\pl_5 +\ka \psi_{5+}\fr{\pl }{\pl \te }+\cdots $
%and in a flat limit coincides with $\pl_5$. The role of '$\ka $-term'
%containing $\psi_{5+}$ will be commented below.
One should note that also compensator hypermultiplets can couple to the 
gauge fields $({\bf V},~{\bf \Si })$, see \cite{kugo01}. Since they have negative kinetic terms, for compensators $(e^{-{\bf V}})^{ab}$ should simply be replaced by $-(e^{-{\bf V}})^{ab}$.
As usual, the exponent of the first term  in eq.(\ref{hypAct}) 
completes  4D derivatives promoting them to 
derivatives covariant under the gauge transformations. In the same way, an additional coupling to the 4D \emph{Weyl} supermultiplet should be included, in order to covariantize the 4D derivatives of the hypermultiplets in respect to the superconformal symmetries. This can however be bypassed by using the (4D) superconformal-invariant D and F-term action formulas of \cite{kak,ueh}, see section \ref{sec:Weylmultiplets}. The superconformal covariant derivatives in the fifth direction, which appear in the kinetic terms, arise in part from the F-term coupling $[({\bf H}\ep )_a (\hat{\pl_y}-{\bf \Si }^{ab}){\bf H}_b]_F$. While ${\bf \Si }^{ab}$ takes care of the gauge invariance, $\hat{\pl_y}$ induces some of the pieces of the superconformal covariant derivatives of $\mcA$ and $\zeta$. The remaing terms are due to the coupling of ${\mathbb W}_y$ to the hypermultiplets in the D-term coupling $[{\mathbb W}_y{\bf H}^{\dagger }{\bf H}]_D$. 
%Note that to obtain the concrete form of the superoperator $\hat\partial_5$ one could start from first principles using the supersymmetry transformations given by FKO to obtain it, or just extract it from the way it must act on ${\bf H}_a$ so that the above action coincides with the one in \cite{kugo01a}.      

One can check that all the relevant (non gravitational)  
couplings of FKO \cite{kugo01a} involving hypermultiplets 
are reproduced by expression (\ref{hypAct}).
We consider now the effects of orbifolding the theory,
 where we can distinguish between the two following special cases:

%\vs{0.3cm} 

%{\bf 1.~Gauge field with positive orbifold parity}

\paragraph{1.~Gauge field with positive orbifold parity:}In this case the 4D gauge superfield ${\bf V}$ and its (5D) partner
${\bf \Si}$
transform under the $Z_2$ orbifold parity ($y\to -y$) as
\beq
Z_2:~~~{\bf V}\to {\bf V}~,~~~~{\bf \Si}\to -{\bf \Si}~.
\la{1Wpar}
\eeq
Therefore 4D $U(1)$ gauge invariance is unbroken at the orbifold fixed points.
With ${\bf H}$  components'  parities
\beq
Z_2:~~~H\to H~,~~~~~H^c\to -H^c~,
\la{1Hpar}
\eeq
for an invariant action, we have to choose   
$q_1=q_1=0$, $q_3=1$. In eq.\eqref{hypAct} for this choice
we have $(e^{-2{\bf V}})^{ab}={\rm Diag}(e^{-2{ V}},~e^{2{ V}})^{ab}$
and (\ref{hypAct}) reduces to
\beq
e_{(4)}^{-1}{\cal L}_{+}({\bf H})=
\int d^4\te\,{\mathbb W}_y 2\l H^{\dagger }e^{-gV}H+H^{c\dagger }e^{gV}H^c\r +
\int d^2\te (H^c\hat{\pl_y}H-H\hat{\pl_y}H^c - 2gH^c\Si H)+{\rm h.c.}
\la{hypActpos}
\eeq
In the rigid limit, this expression coincides with the superspace Lagrangian of \cite{ark01, heb01}. 
It is transparently invariant under the orbifold symmetry, with the 
parity prescriptions (\ref{1Wpar}) and (\ref{1Hpar}). %Note that the action of $\hat\partial_y$ on $H$ and $H^c$ appears in a symmetric way. 
The Lagrangian (\ref{hypAct}) however is more general and allows one to consider 
also the case of a gauge field with negative parity.
Notice that eq.(\ref{hypActpos}) is invariant also if we take for 
$H$ negative parity and for $H^c$ positive parity.

%{\bf 2.~Gauge field with negative orbifold parity}

\paragraph{2.~Gauge field with negative orbifold parity:}In this case 
instead of (\ref{1Wpar}) we have
\beq
Z_2:~~~{\bf V}\to -{\bf V}~,~~~~{\bf \Si}\to {\bf \Si}~,
\la{2Wpar}
\eeq
while for the hypermultiplet the orbifold parities are the same 
as before  (eq.(\ref{1Hpar})).
The requirement of invariance enforces now:
$q_3=0, q_1=\cos \hat{\te } , q_2=\sin \hat{\te }$. 
%
%\beq
%(e^{-2{\bf W}_R})^{ab}=\de^{ab}\cosh (2W)-
%(\si_1\cos \te +\si_2\sin \te )^{ab}\sinh (2W)~.
%\la{exp}
%\eeq
%
With this and the superfield redefinition $H\to e^{-{\rm i}\hat{\te }/2}H$,
$H^c\to e^{{\rm i}\hat{\te }/2}H^c$, eq.(\ref{hypAct}) takes the form
$$
e_{(4)}^{-1}{\cal L}_{-}({\bf H})=
\int d^4\te{\mathbb W}_y 2\l (H^{\dagger }H+H^{c\dagger }H^c)\cosh (gV) -
(H^{\dagger }H^c+H^{c\dagger }H)\sinh (gV)\r +
$$
\beq
\int d^2\te \l H^c\hat{\pl_y}H-H\hat{\pl_y}H^c +
g\Si(H^2-H^{c2})\r+{\rm h.c.} ~.
\la{hypActneg}
\eeq
This expression can be also derived from eq.\eqref{hypActpos}: 
With the parity prescription  eq.(\ref{2Wpar}) and with the modified boundary 
conditions for hypermultiplets
$H(-y)=H^c(y)$, $H^c(-y)= H(y)$, eq.(\ref{hypActpos}) is still invariant. 
Introducing the new combinations ${H}^c_+=\fr{1}{\sq{2}}(H+H^c)$ and
${H}_-=\fr{1}{\sq{2}}(H-H^c)$ (with definite positive and negative 
parities respectively) and rewriting (\ref{hypActpos}) in terms of
${H}_-$, ${H}^c_+$, we recover the Lagrangian eq.(\ref{hypActneg}).

We should emphasize that cases besides these two specific ones, 
i.e. with several gauge fields with different orbifold parities,
can be considered. For such more general cases Lagrangian \eqref{hypAct} should be used.

Alternatively we can introduce an odd gauge coupling $G(y)=\ep (y)g$ 
and use the Lagrangian of case {\bf 1}, eq.\eqref{hypActpos}.  
This is the way FKO 
introduce the $U(1)_R$ gauging of SUGRA to obtain supersymmetric 
RS-like  models \cite{ber, kugo01a}.

Let us now comment on the couplings of the hypermultiplets with the \emph{radion} multiplet. As we pointed out in section 3.2, if one sets the auxilary fields $v_{ay}$, $\lambda^{{\mathbb W}_y}$ and $D^{{\mathbb W}_y}$ to zero, one has $2{\mathbb W}_y=T+T^+$. In this case the D-term coupling in eq.\eqref{hypAct} can be rewritten as
\be
   \int d^4\te\, \frac{T+T^+}{2}2{\bf H}^{\dagger }_a\cdot (e^{-g{\bf V}})^{ab}{\bf H}_b.
\ee
This expression has a form similar to the one which in ref.\cite{marti01} describes the couplings between hypermatter and the \emph{radion} multiplet. But it differs in the fact that the components of ${\bf H}$ also include elements of the radion multiplet, see eq.\eqref{hyperdec}. To understand the relevance of these terms let us consider the case where the only hypermultiplet is a compensator which doesn't couple to any gauge multiplet. One can show that from integrating out $\lambda^{{\mathbb W}_y}$ and $D^{{\mathbb W}_y}$ it follows that $\mcA^2=-2\ka^{-2}$ and $\zeta=0$. This is in fact equivalent to integrating out $\chi'$ and $D'$ in eq.\eqref{L_aux}. If we solve this with $\mcA^{\alpha}_i=\ka^{-1}\delta^{\alpha}_i$ we obtain for the compensator chiral superfields (with the conventions of appendix \ref{sec:spinors})
\be\label{eq:hypercomp1}
            H=\ka^{-1}+\theta^2\left[ \left(i+\frac{W^0_5}{\alpha}\right)\mcF^{2*}_1+\ka^{-1}F_T\right],
\ee
\be\label{eq:hypercomp2}
            H^c=\theta^2 \left(i+\frac{W^0_5}{\alpha}\right)\mcF^{1*}_1,
\ee
where $e^5_y F_T$ is the F-component of $T$. One gets in this way the following
Lagrangian
\be\label{eq:nonflatF_T}
               -\int d^4\te\, \frac{T+T^+}{2}2\left( H^{\dagger }H+H^c H^{c\dagger}\right)=e^5_y\left [\left(1+\frac{(W^0_5)^2}{\alpha^2}\right)\mcF^2+2M_5^3|F_T|^2\right],
\ee
where we used that for compensators $\mcF^2=\mcF^{\alpha}_i d^{\beta}_{\alpha}\mcF^i_{\beta}=-\sum|\mcF^{\alpha}_i|^2$. Eq.\eqref{eq:nonflatF_T} makes evident that due to the breaking of the SU(2)$_R$ by the VEV of $\mcA^{\alpha}_i$, the F-term of the radion superfield, $F_T=-i\ka(V_5^1+iV_5^2)$, is not a flat direction. This means that it cannot be used to induce supersymmetry breaking in the way discussed in \cite{chacko00} (see also \cite{marti01},\cite{ger1,ger2}), at least in the minimal (\emph{ungauged}) version we just described. There is, however, the possibility of extending this by coupling the compensator hypermultiplet to a 5D U(1) vector multiplet \cite{kugo01} with orbifold parity $\Pi(V_R)=-\Pi(\Sigma_R)=-1$. It turns out that in this case the above potential for $F_T$ becomes
\be
         -e^5_y 2M^3_5\left|F_T-ge^{i\alpha}W_{R5}\right|^2,
\ee  
where $\alpha$ is an arbitrary (constant) phase. One sees that if $F_T$ is shifted as $F_T\to F_T+ge^{i\alpha}W_{R5}$, all fields that transform under the SU(2)$_R$ will now couple to the Wilson-line $W_{R5}$ \cite{kugo01}. Clearly, a non-vanishing VEV of $W_{R5}$ will thus lead to supersymmetry breaking. We can now couple ${\mathbb V}_R$ to a tensor multiplet to constrain $W_{R5}$ to be a constant (up to a gauge transformation). This type of SUSY breaking is equivalent to the one discussed in \cite{ger1,ger2}.

\section{The Gauging of 5D SUGRA and the RS Model}\label{sec:gaugesugra}

The purpose of this section is to show how the gauging of 5D \emph{orbifolded} SUGRA, necessary to obtain (local) supersymmetric generalizations of the RS model, can be performed in a very economic way, without having to resort to the 4-form mechanism of \cite{ber} and \cite{kugo01a}. In particular, we consider the case of a single compensator and no physical bulk hypermultiplets. The gauging of 5D SUGRA proceeds by the coupling of the compensating hypermultiplet to a combination ${\mathbb V}\equiv V_I{\mathbb V}^I$ of 5D vector multiplets ${\mathbb V}^I=(\Sigma^I,~V^I)$ with orbifold parities $\Pi (V^I)=-1$. As pointed out in the previous section, if the gauge coupling is odd, $G(y)=g\epsilon (y)$, then we must use the Lagrangian of case {\bf 1}, eq.\eqref{hypActpos}:
\be\label{eq:warpedlagran}
             e_{(4)}^{-1}{\cal L}=
-\int d^4\te\,{\mathbb W}_y 2\l H^{\dagger }e^{-G (y) V}H+H^{c\dagger }e^{G (y) V}H^c\r -
\int d^2\te (H^c\hat{\pl_y}H-H\hat{\pl_y}H^c - 2G (y)H^c\Si H)+{\rm h.c.},
\ee
where $V\equiv V_I V^I$, $\Sigma\equiv V_I \Sigma^I$. 

We assume in the following that the metric is
\be
                 ds^2=e^{2\sigma(y)}dx_{\mu}dx^{\mu}-(e^5_y)^2dy^2.
\ee
The most straightforward way of taking into account the $y$-dependence of this metric is to perform a Weyl rescaling of all fields such that the 4D metric becomes flat, and then use eq.\eqref{eq:warpedlagran} for the rescaled fields. For the vielbein this means
\be
                  e^a_{\mu}\to e^{-\sigma}e^a_{\mu},
\ee
while for the compensating hypermultiplet we have (see eqs.\eqref{eq:hypercomp1},\eqref{eq:hypercomp2}),
\be
                 H=e^{3\sigma/2}\kappa^{-1}-\theta^2 e^{5\sigma/2}F^{*},\quad
H^c=-\theta^2 e^{5\sigma/2}F^{c*}. \ee
For the remaining superfields $V,~\Sigma,~{\mathbb W}_y$ we get
\be
              \Sigma=\left(\tfrac{1}{2}(e^5_y M-iA_y),\cdots\right),\quad V=e^{\sigma}\left(A_{\mu},~e^{\sigma/2}2\Omega_+,~e^{\sigma} D\right), \quad {\mathbb W}_y=(e^{-\sigma}e^5_y,\cdots),
\ee
where $M=V_I M^I$, $A_{\mu}=V_I W^I_{\mu}$. (For a more detailed analysis see appendix \ref{sec:warped}.)

The $F$-term Lagrangian becomes
\be
            \mcL_F\supset e^{4\sigma} -F^c\kappa^{-1}\left[ -3\partial_y
\sigma+ G(y)(e^5_y M-iA_y)\right]+h.c., \ee
while the  $D$-term Lagrangian is
\be
           \mcL_D\supset-e^{4\sigma}e^5_y\, 2\left[
|F|^2+|F^c|^2-\tfrac{1}{2}\kappa^{-2} G(y) D - \kappa^{-1}(F_T F^*+h.c.)
\right], \ee where again $D=V_I D^I$. One can now integrate out the auxiliary
fields $F$ and $F^c$: \be
               F=\kappa^{-1}F_T,\quad F^c=-\frac{\kappa^{-1}}{2e^5_y}\left[
G(y)(e^5_y M+iA_y)-3\partial_y \sigma \right], \ee
to obtain
\be
              \mcL_H =e^{4\sigma}e^5_y M_5^3 \left(2|F_T|^2 +G(y)D+\frac{1}{2}g^2 M^2-3(\partial_5\sigma)G(y)M+\frac{9}{2}(\partial_5\sigma)^2  \right)+\cdots 
\ee
%ere, the first two terms contribute to the scalar potential, in particular the second one gives rise to the brane potentials. 
To obtain the additional contributions to the scalar potential of gauged SUGRA which appear from the vector sector, one has to consider the following terms (see eq.\eqref{eq:vecDterms}):
\be
               \mcL_V\supset-e^{4\sigma}e^5_y \left[\fr{1}{4}\mcN_{IJ}D^I D^J +\frac{1}{2}\mcN_{IJ}D^I \partial_5 M^J+(\partial_5\sigma)\mcN_I D^I \right].
\ee
It is now straightforward to integrate out $D^I$ to obtain
\be\begin{split}\label{eq:RSgeneral1}
              e^{-1}\mcL_{H+V}\supset & -\frac{1}{4}\mcN_{IJ}\partial_a M^I\partial^a M^J+ 6M^3_5 (\partial_5\sigma)^2\\
                                      &\hspace{48pt} +M_5^3 \left(\frac{1}{2}g^2 M^2 + M_5^3 g^2 V_I V_J \mcN^{IJ} + e^y_5 (\partial_y \epsilon (y))gM \right),
\end{split}\ee
where $\mcN^{IJ}$ is the inverse matrix of $\mcN_{IJ}$. 

To put this in a better known form, let us introduce $n$ scalars, $\phi^i$, to parameterise the \emph{very special} manifold defined by $\mcN(M)=\kappa^{-2}$. The $M^I$ ($I=0,\dots,n$) are now functions of the $\phi^i$ ($i=1,\dots,n$). We will use the following definitions:
\be
             g_{ij}(\phi)\equiv -\frac{1}{2}\mcN_{IJ}\frac{\partial M^I}{\partial\phi^i}\frac{\partial M^J}{\partial\phi^j},\quad \mcW(\phi) \equiv M_5^3 g V_I M^I(\phi),
\ee
where $g_{ij}(\phi)$ is the sigma-model metric and $\mcW(\phi)$ the superpotential. We can rewrite eq.\eqref{eq:RSgeneral1} as
\be\label{eq:warplagrfin}
         e^{-1}\mcL=\frac{1}{2} g_{ij}(\phi)\partial_a\phi^i\partial^a\phi^j + 6M^3_5 (\partial_5\sigma)^2 -V_B(\phi)+2e^y_5 [\delta(y)-\delta(y-\pi R)]\,\mcW(\phi),   
\ee
where the bulk potential is now given as
\be
            V_B(\phi)=\frac{1}{2}g^{ij}\mcW_i\mcW_j-\frac{2}{3M_5^3}\mcW^2,
\ee
with $\mcW_i=\partial\mcW/\partial\phi^i$ and $g^{ij}$ being the inverse of the metric $g_{ij}$.\footnote{To obtain this result we used the following relation $g^{ij}\mcN_{I,i}\mcN_{J,j}=4 a_{IJ}-2\mcN_I\mcN_J/3\mcN$.} These results agree precisely with the ones obtained with the 4-form mechanism \cite{ber},\cite{kugo01a}. Note how supersymmetry relates the brane potentials to the bulk one. In particular, if we take no physical (bulk) vector multiplets, then we have $\mcW_i=0$ and thus the brane tensions, $\tau=\pm 2\mcW$, are tuned with the bulk cosmological constant, with $\tau=\pm \sqrt{-6M^3_5 V_B}$, as in the Randall-Sundrum model \cite{ran1}. The equation of motion for the warp-factor $\sigma(y)$ then follows from the Lagrangian, eq.\eqref{eq:warplagrfin}, giving the solution $\sigma(y)=\pm\sqrt{-V_B/6M^3_5}|y|$.

Closing this section, let us note that from the conditions $F^c=D^I=0$ one gets
$$
                 \partial_y\si=e^5_y \frac{\ep(y)}{3M^3_5}\mcW,
$$
\be
  g_{ij}\pl_5\phi^j=-  \ep(y)\fr{\pl\mcW}{\pl\phi^i},                 
\ee
which are nothing but the BPS conditions (to be compared with expressions of \cite{ber}), obtained here in a rather simple way.

\section{4D Weyl Multiplet and Couplings with Matter}
\label{sec:Weylmultiplets}

In our formulations we have used superspace actions, which produce $F$ and
$D$ terms of various operators upon integrating over $d^2\te $ and
$d^4\te $ respectively. Obviously, this does not account for the couplings of 
matter (i.e. gauge multiplets and hypermultiplets) with 4D SUGRA. In order to
achieve this, one should first obtain the 4D Weyl multiplet induced from the 5D Weyl super multiplet. The components of the former are \cite{kugo02}
\beq
e^{(4)a}_{\mu }=e^{a}_{\mu }~,~~\psi^{(4)}_{\mu }=\psi_{\mu +}~,~~
b^{(4)}_{\mu }=b_{\mu }~,~~
A^{(4)}_{\mu }=\fr{4}{3}(V_{\mu }^3+v_{\mu 5})~.
%
%
%\phi^{(4)}_{\mu }=
%\phi_{\mu -}-\ga_5\ga^av_{a5}\psi_{\mu +}+\De \phi_{\mu }~,~~~~
%f^{(4)a}_{\mu }=f^{a}_{\mu }-\bar \psi_{\mu +}\De \phi^a+\De f^{a}_{\mu }~.
\la{4Dweyl}
\eeq
%with
%\beq
%\De \phi_{\mu }\equiv \fr{{\rm i}}{2}\ga_5\hat{R}_{\mu 5}(Q)_{-}~,~~~
%\De f^{a}_{\mu }\equiv -\fr{1}{6}\ep^{abc}_{\mu }\l \hat{D}_bv_{c5}+
%\fr{1}{2}\hat{R}^3_{bc}(V)\r +\fr{1}{4}\hat{R}_{\mu 5}^{a5}(M)~.
%\la{deltas}
%\eeq
%
Note that the composition of the 4D Weyl multiplet and of the radion
supermultiplet are independent also with respect to the auxiliary 
fields entering in both multiplets. This is necessary since the radion
should transform independently of the 4D Weyl multiplet \cite{lut}.
Note also that the definition eq.(\ref{4Dweyl}) of the 4D Weyl multiplet 
(as well as that of compensators) is unique and does not
allow any addition as emphasized by the authors of \cite{kugo02} 
(this is not the case for the approach of ref. \cite{zucker}).

We could now assemble the components of the 4D Weyl multiplet into a superfield, 
which would than be coupled to the other superfields. This is well known but
its usefullness is not clear if one has in mind that one would have to
calculate in some background (e.g. AdS). There is, however, an alternative
formalism which gives the very same couplings we are searching for
\cite{kak,ueh,kkl}. It consists of replacing the $d^2\theta$ and $d^4\theta$
integrations in the Lagrangian by $F$-terms and $D$-terms invariant under the
4D conformal SUGRA. Knowing this, the $F$-term of a chiral operator of weights
$(3,3)$,  ${\cal C}=[\mcA,~\chi_R,~\mcF]^{\cal C}$ (which in general is a
composite chiral superfield), should be  understood as follows
\cite{kak,ueh,kkl} \beq [{\cal C}]_{\bf F}=e^{(4)}\l \int \hs{-0.3mm}d^2\te
{\cal C}+{\rm h.c.}\r - e^{(4)}\left [ i\ka {\bar \psi}^{(4)}\cdot \ga
\chi^{\cal C}_R+ 2\ka^2{\bar\psi}^{(4)}_{a}\ga^{ab}\psi^{(4)}_{Lb}\mcA+{\rm
h.c.}\right ]~. \la{sugraF} \eeq In the same way, the $D$-term of a real
general operator with weights $(2,0)$, ${\cal V}=[C, \zeta, H, K, B_{a}, \lam
, D]^{\cal V}$ (which in general is a composite real multiplet), must be
understood as follows $$ [{\cal V}]_{\bf D}=e^{(4)}  \int \hs{-0.3mm}d^4\te
\,2{\cal V}+ e^{(4)}\left [ -\ka \bar \psi^{(4)}\cdot \ga \ga_5\lam^{\cal V}+
\ka \frac{2}{3}i\bar\zeta^{\mcV}\ga_5 \ga^{\mu\nu}\mcD_{\mu}\psi^{(4)}_{\nu}+
\fr{1}{3}C^{\cal V}({\mathcal R}^{(4)}+4{\rm i}\ka^2\bar \psi^{(4)}_{\mu
}\ga^{\mu\nu\lam}\mcD_{\nu}\psi^{(4)}_{\lam})   \right.  $$
\beq
\left. +{\rm i}\ka^2\ep^{abcd}\bar \psi^{(4)}_{a }
\ga_{b }\psi^{(4)}_{c}(B^{\cal V}_{d}
-\ka\bar \psi^{(4)}_{d}\zeta^{\cal V})\right ].
\la{sugraD}
\eeq
(See \cite{kugo02} for the definitions of 
${\mathcal R}^{(4)}$, $\mcD_{\mu}\psi^{(4)}_{\lambda}$ etc.)
As we see, the first terms in (\ref{sugraF}), (\ref{sugraD}) are
the contributions which we get upon integration over the $\te $ 
superspace coordinates. The remaining terms give interactions with the 
components of the 4D Weyl multiplet. Therefore, for calculating all relevant 
couplings, in our superspace actions we should make the replacements
\beq
e^{(4)}\int \hs{-1mm}d^2\te (\cdots )+{\rm h.c.}\to [\cdots ]_{\bf F}~,~~~~~~~
e^{(4)}\int \hs{-1mm}d^4\te (\cdots )\to \frac{1}{2}[\cdots ]_{\bf D}~,
\la{sugraRepl}
\eeq
and then use expressions (\ref{sugraF}) and (\ref{sugraD}) for 
their evaluation. Note also that when applying these expressions to composite multiplets, one should use the formulas for products of general multiplets of 4D conformal SUGRA given in \cite{kugo02,kak} to ensure that all 4D derivatives are covariant in respect to 4D conformal SUGRA.

As an illustration of the relevance of this procedure let us look at the D-terms present in our actions:
\be
               \frac{1}{2}\left[ {\mathbb W}_y(2P(\mcV_5)+{\bf H}^{+}\cdot{\bf H})\right]_{\bf D}\supset e^{(4)}e^5_y\left(\frac{1}{8}D(2\mcN+\mcA^2)-\frac{1}{6}(\mcN-\mcA^2){\mathcal R}^{(4)}\right).
\ee
From a variation of the auxiliary field $D$ it follows that $\mcA^2=-2\mcN$ and therefore $\mcN-\mcA^2=3\mcN$. In this way one gets the correct coupling between the Ricci scalar and $\mcN$,
\be
                -\frac{1}{2}\mcN {\mathcal R}^{(4)},
\ee
which becomes canonical after the conformal gauge fixing $\mcN=\ka^{-2}$.

%Besides $C^{\cal V}$-contribution in (\ref{sugraD}), the five dimensional
%curvature gets contribution from the five dimensional bulk term 
%$-\fr{1}{2\ka^2 }R^{(5)}$. After Weyl transformation of vierbein
%$\tl{e}^{{(4)}a}_{\mu }=
%e^{(4)a}_{\mu }(1-\fr{2\ka^2}{3}\sum C^{\cal V})^{1/2}$
%the effective 4D gravitational constant is now actually constant.

As we mentioned in the previous sections, what is not fully treated in our 
superspace formulation up to now is the odd (with respect 
to orbifold parity) part of the 5D Weyl multiplet. The same is still true 
for the $\pl_y$ derivatives of the 4D Weyl multiplet (though we have sketched
the covariantization of $\pl_y$).
Additional effort is needed in order to embed these degrees of 
freedom into 4D superfields. Once this is done, one should be able to 
account for the full 5D covariance 
as well as for the couplings between the matter multiplets 
and that odd part induced from the 5D Weyl supermultiplet. 
However, this goes beyond the scope of this paper.
%zzz

\section{Conclusions}

We argued that the very powerful formulation of 5D conformal SUGRA of 
FKO can be stenographically written in 4D superfield language and 
we gave explicit expressions for the radion-gauge-hypermultiplet sector.

On our way to this aim we have identified the 
radion chiral superfield discussed in the literature in the decomposition of the 
5D 'graviphoton' gauge supermultiplet of the superconformal approach. In addition we 
found another superfield, in the reduction of the 5D Weyl multiplet, containing 
elements of the radion multiplet.We then have presented a Chern-Simons 
type superfield Lagrangian with a prepotential which exactly 
reproduces the gauge part of FKO. It is closely related to expressions 
for global 5D SUSY given in the literature and reduces to it 
for $M_{5}\to \infty $. 
It also contains the interaction with the radion superfield. 
Different from 
what one is used to, also the auxiliary components of the other 
superfields contain the components of the radion multiplet. 
It is a strong advantage 
of the present approach that one can easily write down its interactions. 
The Lagrangian proposed in ref.\cite{marti01} is contained in our expressions.
Our construction allows for a simple derivation of FI-terms within 5D orbifold 
SUGRA.

The 5D hypermultiplet action is also formulated in superspace and reduces 
to the known case of rigid SUSY. We have discussed the coupling of the 
matter fields to gauge superfields and their associated chiral fields, 
considering both orbifold parities of the gauge field. Since also the 
compensators are hypermultiplets, in this approach they are described simultaneously 
in the same superformalism.

The remaining gravitational interaction can be formulated introducing 
the 4D Weyl supermultiplet, with even orbifold parity, originally inherited from the 5D Weyl supermultiplet. This interaction 
can be formulated in well known superfield language if required. What is 
left is the odd orbifold parity part of the 5D Weyl supermultiplet which should
also fit into a $\mcN=1$, 4D formulation. We have not spelled out that completely. However, we have presented the structure
of a supercovariant derivative ${\hat\partial}_5$ which accounts for the hypermultiplet
 interaction with the odd part of the Weyl multiplet. Inclusion of the latter 
insures 5D covariance. This would
automatically account for the second SUSY.
Finally all terms in the FKO Lagrangian(s) should be 
reproduced. If this goes through, we would have the full theory without 
explicit use of the second (parity odd) supersymmetry. In present modeling
\cite{rat} the 
not well understood role of this second SUSY restricts the reliability, e.g. 
restricts the radion superfield to its zero mode part.

As far as the $\mcN=1$ SUSY (which survives at 4D after orbifold
compactification) is concerned, it can be broken by a non zero VEV 
of the auxiliary component $F_{T }$ of the radion supermultiplet. It
provides a soft mass for the gauginos [due to the first term in 
(\ref{eq:vecsuperlagr})], as
well as a soft mass squared for the lowest scalar components of the
hypermultiplets [see couplings in (\ref{hypAct})]. 
Another possibility is the SUSY breaking through the gaugino condensation.
These mechanisms should lead to phenomenologically satisfactory scenarios 
\cite{lut, ger2, rat}.

To obtain warped solutions we have considered the gauging of the U(1)$_R$
in orbifold SUGRA. The superfield construction offers a simple way of 
obtaining the tuned bulk and brane potentials, as well as the BPS equations.

It is a nice property of 5D conformal SUGRA that it smoothly transfers 
to 4D conformal SUGRA on the brane boundaries. Thus fields based solely 
on the brane(s) can be easily coupled respecting the symmetries. This has 
already been demonstrated in FKO work. 

Odd-odd (under orbifold parity) fermionic terms arise already from 
additional terms in the auxiliary components of the newly defined 
superfields, but they also appear spelling out the interaction with the
odd part of the Weyl supermultiplet. Their effect in the brane 
interaction should be easier to study in this approach. Altogether we 
hope to simplify model 
building in the context of 5D conformal 
SUGRA allowing for transparent formulations.

%zzz

\vs{0.5cm}
 
\hs{-0.7cm}{\bf Acknowledgments}
 
\vs{0.2cm}  
\hs{-0.7cm}We would like to thank T. Kugo for a very helpful 
correspondence at the beginning of this work. M.G.S. also thanks 
N. Dragon, T. Gregoire, Z. Lalak, H.P. Nilles, R. Rattazzi and C. Scrucca
for useful discussions and the CERN theory division for hospitality. The
research of F.P.C. is supported by Funda\c c\~ ao para a Ci\^ encia e a
Tecnologia (grant SFRH/BD/4973/2001).

%%%%%%%%%%%%%%%%%%%%%%%%%%%%%%%%%%%%%%%%%%%%%%%%%%%%%%%%%%%%%%%%%%%%%%%%%%%%%%%%%%%%%%%%%%%%%%%%%%%%%  APPENDICES  %%%%%%%%%%%%%%%%%%%%%%%%%%%%%%%%%%%%%%%%%%%%%%%%%%%%%%%%%%%%%%%%%%%%%%%%%%%%%%%%%%%%%%%%%%%%%%%%%%%%%%%%%%%%%%%%%%%%%%%%%%

\appendix

\renewcommand{\theequation}{A.\arabic{equation}}\setcounter{equation}{0}
\section{Field content and component supergravity action}\label{sec:actionsFKO}

The field content of the introduced supermultiplets and orbifold parity
assignments are given in Table 1.

\begin{table}[!h]
\begin{center}
\begin{tabular}{|c|c|}
\hline 
$Z_2$ parity & Field content\\
\hline\hline
\multicolumn{2}{|c|}{Weyl multiplet}\\
\hline
+ & $e^a_{\mu},~e^5_y,~\psi_{\mu+},~\psi_{y-},~b_{\mu},~V_{\mu}^3,~V_y^{1,2},~v^{5a},~\chi_+,~D $\\
\hline
-& $e^5_{\mu},~e^a_y,~\psi_{\mu-},~\psi_{y+},~b_{y},~V_{y}^3,~V_{\mu}^{1,2},~v^{ab},~\chi_-$ \\
\hline\hline
\multicolumn{2}{|c|}{Vector multiplet}\\
\hline
$\Pi_V$ & $M,~W_y,~\Omega_-,~Y^{1,2}$ \\
\hline
$-\Pi_V$ & $W_{\mu},~\Omega_+,~Y^3$ \\
\hline\hline
\multicolumn{2}{|c|}{Hypermultiplet}\\
\hline
$\Pi_{\hat\alpha}$ & $\mcA^{2{\hat\alpha}-1}_1,~\mcA^{2{\hat\alpha}}_2,~\zeta^{\hat\alpha}_-,~\mcF^{2{\hat\alpha}-1}_1,~\mcF^{2{\hat\alpha}}_2$\\
\hline
$-\Pi_{\hat\alpha}$ & $\mcA^{2{\hat\alpha}-1}_2,~\mcA^{2{\hat\alpha}}_1,~\zeta^{\hat\alpha}_+,~\mcF^{2{\hat\alpha}-1}_2,~\mcF^{2{\hat\alpha}}_1$\\
\hline
\end{tabular}
\end{center}
\caption{Field Content and $Z_2$ Parities}
\end{table}

We present now the Lagrangians obtained in \cite{kugo01,kugo01a} with explicit powers of the 5D Planck mass $M_5$. Note that here the conformal symmetries are already fixed. The gravity/vector part is, with $\ka\equiv M_5^{-3/2}$,
{\setlength\arraycolsep{2pt}\begin{eqnarray}
 e^{-1}{\mathcal L}_{GV} & = & -\frac{1}{2\ka^2}{\mathcal R}(\omega)-2i{\bar \psi}_{\mu}\gamma^{\mu\nu\rho}\nabla_{\nu}\psi_{\rho}+\ka^2{\bar \psi}_a\psi_b({\bar \psi}_c\gamma^{abcd}\psi_d+{\bar \psi}^a\psi^b) \nonumber\\
                                             & & {}+{\mathcal N}_I(g[{\bar\Omega},\Omega]^I+\ka^2\frac{i}{4}F_{ab}(W)^I{\bar \psi}_c\gamma^{abcd}\psi_d)+a_{IJ}f_1^{IJ}(W,\Omega,M)\nonumber\\
                                             & & {}-{\mathcal N}_{IJK}f_2^{IJK}(W,\Omega,M)+\frac{\ka^2}{8}\left(2{\bar \psi}_a\psi_b+{\bar\zeta}^{{\bar \alpha}}\gamma_{ab}\zeta_{\alpha}+a_{IJ}{\bar\Omega}^I\gamma_{ab}\Omega^J   \right)^2\nonumber\\
                                             & & {}+\ka^2\frac{i}{4}{\mathcal N}_I F^{ab}(W)^I(2{\bar \psi}_a\psi_b+{\bar\zeta}^{{\bar \alpha}}\gamma_{ab}\zeta_{\alpha}+a_{IJ}{\bar\Omega}^I\gamma_{ab}\Omega^J)\nonumber\\
                                             & & {}+\ka^2({\mathcal A}^{{\bar \alpha}i}\nabla_a{\mathcal A}^j_{\alpha}+ia_{IJ}{\bar\Omega}^{Ii}\gamma_{a}\Omega^{Jj})^2,
\la{lagFKOrev}
\end{eqnarray}}
where
{\setlength\arraycolsep{2pt}\begin{eqnarray}
                               f_1^{IJ} & = & -\frac{1}{4}F(W)^I\cdot F(W)^J+\frac{1}{2}\nabla M^I\cdot\nabla M^J+2i{\bar\Omega}^I\gamma\cdot\nabla\Omega^J\nonumber\\
                                        & & {}+i\ka{\bar \psi}_a(\gamma\cdot F(W)-2\gamma\cdot\nabla M)^I\gamma^a\Omega^J\nonumber\\
                                        & & {}-2\ka^2\{({\bar\Omega}^I\gamma^a\gamma^{bc}\psi_a)({\bar \psi}_b\gamma_c\Omega^J)-({\bar\Omega}^I\gamma^a\gamma^{b}\psi_a)({\bar \psi}_b\Omega^J)\},
\end{eqnarray}}
{\setlength\arraycolsep{2pt}\begin{eqnarray}
                               f_2^{IJK} & = & -\frac{i}{4}{\bar\Omega}^I\gamma\cdot F(W)^J\Omega^K+\frac{2\ka}{3}({\bar\Omega}^I\gamma^{ab}\Omega^J)({\bar \psi}_a\gamma_b\Omega^K)\nonumber\\
                                         & & {}+\frac{2\ka}{3}({\bar \psi}^i_a\gamma^a\Omega^{Ij})({\bar\Omega}^J_{(i}\Omega^K_{j)}),
\end{eqnarray}}
and ${\mathcal N}_I$, ${\mathcal N}_{IJ}$, ${\mathcal N}_{IJK}$, $a_{IJ}$ are functions of the scalar components of the vector multiplets, $M_I$. These functions are obtained through differentiation of the \emph{norm function} ${\mathcal N}(M_I)$, a homogeneous cubic function of the $M_I$ which caracterizes the vector part of the system:
\be
                      {\mathcal N}_I\equiv \frac{\partial {\mathcal N}}{\partial M^I},\quad  {\mathcal N}_{IJ}\equiv \frac{\partial^2 {\mathcal N}}{\partial M^I\partial M^J},\quad{\mathcal N}_{IJK}\equiv \frac{\partial^3 {\mathcal N}}{\partial M^I\partial M^J\partial M^K},
\ee
\be
                   a_{IJ}=-\frac{1}{2\ka^2}\frac{\partial^2}{\partial M^I\partial M^J}\ln (\ka^2{\mathcal N}).
\ee

There is also a Chern-Simons Lagrangian,
{\setlength\arraycolsep{2pt}\begin{eqnarray}
                        e^{-1}{\mathcal L}_{C-S} & = & \frac{\ka}{8}c_{IJK}\epsilon^{\lambda\mu\nu\rho\sigma}W^I_{\lambda}\left(F_{\mu\nu}^J(W)F_{\rho\sigma}^K(W)+\frac{i}{2}g[W_{\mu},W_{\nu}]^J F_{\rho\sigma}^K(W)\nonumber\right.\\
                                                 & & \qquad\qquad\qquad\qquad\qquad\left. -\frac{g^2}{10}[W_{\mu},W_{\nu}]^J[W_{\rho},W_{\sigma}]^K\right),
\end{eqnarray}}
where $c_{IJK}={\mathcal N}_{IJK}/6$, and a hypermultiplet Lagrangian,
{\setlength\arraycolsep{2pt}\begin{eqnarray}
                       e^{-1}{\mathcal L}_{hyper} & = & \nabla^a {\mathcal A}_i^{\bar\alpha}\nabla_a{\mathcal A}^i_{\alpha}-2i{\bar\zeta}^{\bar\alpha}(\gamma\cdot\nabla+igM)\zeta_{\alpha}-{\mathcal A}_i^{\bar\alpha}(gM)^{2~~\beta}_{~\alpha}{\mathcal A}^i_{\beta}\nonumber\\
                                                  & & {}-4i\kappa{\bar\psi}^i_a\gamma^b\gamma^a\zeta_{\alpha}\nabla_b{\mathcal A}_i^{\bar\alpha}-2i\kappa^2{\bar\psi}_a^{(i}\gamma^{abc}\psi^{j)}_c{\mathcal A}_j^{\bar\alpha}\nabla_b{\mathcal A}_{\alpha i}\nonumber\\
                                     & & {} +{\mathcal A}_i^{\bar\alpha}\left(-8g{\bar\Omega}^i_{\alpha\beta}\,\zeta^{\beta}+4\kappa g{\bar\psi}_a^i\gamma^a M_{\alpha\beta}\zeta^{\beta} -4\kappa g{\bar\psi}_a^{(i}\gamma^a {\Omega}_{\alpha\beta}^{j)}{\mathcal A}^{\beta}_{j} \right.\nonumber\\
                                     & & \left. +\, 2\kappa^2 g {\bar\psi}_a^{(i}\gamma^{ab}\psi^{j)}_b M_{\alpha\beta}{\mathcal A}^{\beta}_{j}\right)+\kappa^2{\bar\psi}_a\gamma_b\psi_c{\bar\zeta}^{\bar\alpha}\gamma^{abc}\zeta_{\alpha}\nonumber\\
                                     & &{}-\frac{\kappa^2}{2}{\bar\psi}^a\gamma^{bc}\psi_a{\bar\zeta}^{\bar\alpha}\gamma_{bc}\zeta_{\alpha}.
\end{eqnarray}}
Here we assumed that the hyperfields transform under the gauge group $G$ ($I\neq 0$) as 
\be
                 \delta{\mathcal A}_i^{\alpha}=i\omega^I(t_I)^{\alpha}_{\beta}{\mathcal A}_i^{\beta}.
\ee

Finally we present the so-called auxiliary Lagrangian, which appart from the $Y$-terms vanishes on-shell: 
{\setlength\arraycolsep{2pt}\begin{eqnarray}\label{L_aux}
                 e^{-1}{\mathcal L}_{aux} & = & D'(\kappa^2{\mathcal A}^2+2)-8i\kappa {\bar\chi}'^i{\mathcal A}_i^{\bar\alpha}\zeta_{\alpha}+(Y\textup{-terms})+2(v-{\tilde v})^{ab}(v-{\tilde v})_{ab}\nonumber\\
                                   &   +&(V_{\mu}-{\tilde V}_{\mu})^{ij}(V^{\mu}-{\tilde V}^{\mu})_{ij}+
\left(1-\frac{(W^0_a)^2}{\alpha^2}\right)({\mathcal F}_i^{\bar\alpha}-{\tilde{\mathcal F}}_i^{\bar\alpha})({\mathcal F}^i_{\alpha}-{\tilde{\mathcal F}}^i_{\alpha}),\nonumber\\
                      & & 
\end{eqnarray}}
where the $Y$-terms are given by
\be
              -\frac{1}{2}{\mathcal N}_{IJ}Y^I_{ij}Y^{Jij}+Y^I_{ij}\left[2i{\mathcal A}_{\alpha}^{(i}(gt_I)^{{\bar\alpha}\beta}{\mathcal A}^{j)}_{\beta}+i{\mathcal N}_{IJK}{\bar\Omega}^{Ji}\Omega^{Kj}\right],
\ee
and ${\tilde v}_{ab}$, $\tilde{\mathcal F}^i_{\alpha}$, ${\tilde V}_{\mu}^{ij}$, are combinations of non-auxiliary fields \cite{kugo01}:
\be
                {\tilde V}_{a}^{ij}=-\frac{\kappa}{2}(2{\mathcal A}^{{\bar\alpha}(i}\nabla_a  {\mathcal A}_{\bar\alpha}^{j)}-i{\mathcal N}_{IJ}{\bar\Omega}^I\gamma_{ab}\Omega^J),  
\ee
\be\label{eq:va_ab_sol}
           {\tilde v}_{ab}=-\frac{\kappa}{4}\left[ \mcN_I F_{ab}^I-i(6{\bar\psi}_a\psi_b+{\bar\zeta}^{\alpha}\gamma_{ab}\zeta_{\alpha}-\frac{1}{2}\mcN_{IJ}{\bar\Omega}^I\gamma_{ab}\Omega^J)\right]
\ee
\be
           {\tilde\mcF}^{\alpha }_i=(gM^0t_0)^{\alpha}_{~\beta}\mcA_i^{\beta}
\ee

%%%%%%%%%%%%%%%%%%%%%%%%%%%%%%%%%%%%%%%%%%%%%%%%%%%%%%%%%%%%%%%%%%%%%%%%%%%%%%%%%%%%%%%%%%%%%%%%%%%%%%%%%%%  THE CONSTRAINTS  %%%%%%%%%%%%%%%%%%%%%%%%%%%%%%%%%%%%%%%%%%%%%%%%%%%%%%%%%%%%%%%%%%%%%%%%%%%%%%%%%%%%%%%%%%%%%%%%%%%%%%%%%%%%%%%

\renewcommand{\theequation}{B.\arabic{equation}}\setcounter{equation}{0}
\section{The constraints}\label{constraints} 

The action, as it stays above, is the result of fixing the superconformal symmetries (${\bf D},~{\bf S}^i,~{\bf K}_a$) with the following constraints on the norm function:
\be
                     {\mathcal N}=\ka^{-2},\quad {\mathcal N}_I\Omega^I=0,\quad {\hat{\mathcal D}}_a {\mathcal N}=0.
\ee 
Consider first the last constraint ${\hat{\mathcal D}}_a {\mathcal N}=0$. To do this note that \cite{kugo02} 
\be
                {\hat{\mathcal D}}_a M^I=(\partial_a-b_a)M^I-ig[W_a,M]^I-2i\kappa{\bar\psi}_a\Omega^I,
\ee
where $b_a$ is the gauge field of dilatation and $\psi_a$ is the gravitino (the gauge field of susy). Now, ${\mathcal N}$ is a gauge invariant, has Weyl weight $w=3$, therefore one has
\be
                 {\hat{\mathcal D}}_a {\mathcal N}=(\partial_a-3b_a){\mathcal N}-2 i\ka {\bar\psi}_a\frac{\partial {\mathcal N}}{\partial M^I}\Omega^I=(\partial_a-3b_a){\mathcal N}-2i\kappa{\bar\psi}_a {\mathcal N}_I\Omega^I. 
\ee 
With the constraints ${\mathcal N}=\kappa^{-2}$ and ${\mathcal N}_I\Omega^I=0$, it follows that 
\be
                    b_a=0.
\ee

To solve the other two constraints one must precise the vector multiplets. Let us consider as an example in addition to the compensator vector multiplet $V^0$ just another  $U(1)$ vector multiplet. In this case the most general norm function is 
\be
                   \ka^{-1}{\mathcal N}=\alpha {M^{0}}^3-\beta\alpha^{\frac{1}{3}} M^0 {M^1}^2-\gamma{M^1}^3.          
\ee
To obtain canonically normalized fields we set $\alpha=(2/3)^{\frac{3}{2}}$ and $\beta=1$. Introducing $\phi_1\equiv M^1/\alpha^{\frac{1}{3}}\ka M^0$, one can rewrite the constraint $\mcN=\ka^{-2}$ as
\be
              \ka^{-1}{\mathcal N}=\alpha{M^0}^3(1-\ka^2\phi_1^2-\gamma\ka^3\phi_1^3)=\ka^{-3}.  
\ee
This is solved by
\be
                    \ka M^0(\phi_1)=\left(3/2\right)^{\frac{1}{2}}\left[1-\ka^2\phi_1^2-\gamma\ka^3\phi_1^3 \right]^{-\frac{1}{3}}\simeq \left(3/2\right)^{\frac{1}{2}}\left[1+\frac{1}{3}\ka^2\phi_1^2+\frac{1}{3}\gamma\kappa^3\phi_1^3+\mcO \left((\ka\phi_1)^4\right)\right].
\ee
Clearly we are assuming that in the vacuum $M^0=\left(3/2\right)^{\frac{1}{2}}\ka^{-1}$ and $M^1=\phi_1=0$, and an expansion in powers of $\ka M^1$ is reasonable.

\renewcommand{\theequation}{C.\arabic{equation}}\setcounter{equation}{0}
\section{Conventions and superfield definitions}\label{sec:spinors}

In this appendix we present some conventions and expressions for the 
superfields in four component Majorana spinors, as well as in two component 
Weyl spinors.

The 5D $\ga $-matrices, satisfying relations $\{\ga^a, \ga^b \}=2\eta^{ab}$ 
with $\eta^{ab}={\rm Diag}(1, -1, -1, -1, -1)^{ab}$, are
\begin{equation}
\begin{array}{cc}
 & {\begin{array}{cc}
~ & 
\end{array}}\\ \vspace{2mm}
\begin{array}{c}
 \\  \\ 
\end{array}\!\!\!\!\! &\ga^0={\left(\begin{array}{cc}
0 ~&{\bf 1} 
\\
{\bf 1}~&0
\end{array}\right)~, }~
\end{array}  
\begin{array}{cc}
 & {\begin{array}{cc}
~ & 
\end{array}}\\ \vspace{2mm}
\begin{array}{c}
 \\  \\ 

\end{array}\!\!\!\!\! &\ga^i={\rm i}{\left(\begin{array}{cc}
0~& \si^i
\\
-\si^i ~& 0
\end{array}\right) }~,~~
\end{array}  
\hs{-1cm}
\begin{array}{ccc}
& {\begin{array}{ccc}
 &   
\end{array}}\\ \vspace{2mm}
\begin{array}{c}
  \\  \\
\end{array} &\ga^4={\rm i}{\left(\begin{array}{ccc}
-{\bf 1}~ & 0
\\
0~ & {\bf 1}
\end{array}\right)~,~~~~i=1, 2, 3~.
}
\end{array}
\label{5Dgamas}
\end{equation}
The charge conjugation matrix $C_5$, satisfying 
$C_5^T=-C_5$, $C_5^{\dagger }C_5={\bf 1}$, $C_5\ga_aC_5^{-1}=\ga^T$, in
(\ref{5Dgamas}) representation is $C_5={\rm Diag}(\ep , \ep)$. 5D spinors
$\psi^i$, being doublets of $SU(2)_R$, can be represented through two 
component Weyl spinors
\beq
\psi^i=\l (\chi^i)_{\al},~~ \bar{(\xi^i)}^{\hs{0.2mm}\dot{\al }}\r^T~,
\la{5DspinotSU2}
\eeq
where $\al, \dot{\al }=1,2$. The indices are lowered and raised by $SU(2)$
antisymmetric tensors:
$\psi_i=\psi^j\ep_{ji}$, $\chi_{\al}=\ep_{\al \bt }\chi^{\bt }$,
$\chi^{\al }=\chi_{\bt }\ep^{\bt \al }$ (similar for dotted indices),
$\ep^{12}=\ep_{12}=1$. Upon imposing the $SU(2)$ Majorana condition 
$\ov{\psi }^i\equiv (\psi_i)^{\dagger }\ga^0=(\psi^i)^TC_5$, it is easy to 
check out that
$$
\psi^1=\l (\chi^1)_{\al},~~ \bar{(\chi^2)}^{\hs{0.2mm}\dot{\al }}\r^T~,~~~
\psi^2=\l (\chi^2)_{\al},~~ -\bar{(\chi^1)}^{\hs{0.2mm}\dot{\al }}\r^T~,
$$
\beq 
\ov{\psi }^{\hs{0.3mm}1}=\l (\chi^1)^{\hs{0.2mm}\al},~~ 
-\bar{(\chi^2)}_{\dot{\al }}\r~,~~~
\ov{\psi }^{\hs{0.3mm}2}=\l (\chi^2)^{\hs{0.2mm}\al},~~ 
\bar{(\chi^1)}_{\dot{\al }}\r  ~.
\la{5Dsprepr}
\eeq
We can use this notation for the parameterization of gravitinos. Similarly 
four component gauginos $\Om^i$ can be written,  with $\chi^i $ in
(\ref{5Dsprepr}) replaced by $\om^i $. Treating the hypermultiplet's
fermionic components $(\zeta^{2\hat{\al }-1}, \zeta^{2\hat{\al }})$
-  hyperinos - in analogy with $(\psi^i, \psi^2)$, it is convenient to use the 
following parameterization:
$$
\zeta^{2\hat{\al }-1}\equiv \zeta^1=
\l (\eta^1)_{\al},~~ \bar{(\eta^2)}^{\hs{0.2mm}\dot{\al }}\r^T~,~~~
\zeta^{2\hat{\al }}\equiv \zeta^2=\l (\eta^2)_{\al},~~ 
-\bar{(\eta^1)}^{\hs{0.2mm}\dot{\al }}\r^T~,
$$
\beq
\ov{\zeta }^{\hs{0.3mm}2\hat{\al }-1}=
\ov{\zeta }^{\hs{0.3mm}1}=\l (\eta^1)^{\hs{0.2mm}\al},~~ 
-\bar{(\eta^2)}_{\dot{\al }}\r~,~~~
\ov{\zeta }^{\hs{0.3mm}2\hat{\al }}=\ov{\zeta }^{\hs{0.2mm}2}=
\l (\eta^2)^{\hs{0.3mm}\al},~~ \bar{(\eta^1)}_{\dot{\al }}\r  ~.
\la{5Dhyperino}
\eeq

In 4D, the charge conjugation matrix $C_4$ satisfy 
$C_4^T=-C_4$, $C_4^{\dagger }C_4={\bf 1}$, $C_4\ga_aC_4^{-1}=-\ga^T_a$
and is given by $C_4={\rm Diag}(\ep, -\ep )$. The 4D Majorana condition 
$\ov{\psi }\equiv \psi^{\dagger }\ga^0=\psi^TC_4$ defines the 4D Majorana
spinor 
$\psi =\l (\lam )_{\al},~~ \bar{(\lam )}^{\hs{0.2mm}\dot{\al }}\r^T$~.
The 4D chirality matrix $\ga_5$ is related with $\ga^4$ by 
$\ga_5=-{\rm i}\ga^4$ and defines left and right projection operators
\beq
{\cal P}_L=\fr{1}{2}(1-\ga_5)~,~~~ {\cal P}_R=\fr{1}{2}(1+\ga_5)~,~~~~
{\rm with}~~~
\psi_L={\cal P}_L\psi ~,~~~ \psi_R={\cal P}_R\psi ~.
\la{projOps}
\eeq

From 5D spinors (\ref{5Dsprepr}) one can build the combinations
\beq
\psi_{+}=\psi^1_R+\psi^2_L=
\l (\chi^2)_{\al},~~ \bar{(\chi^2)}^{\hs{0.2mm}\dot{\al }}\r^T~,~~~
\psi_{-}={\rm i}(\psi^1_L+\psi^2_R)=
{\rm i}\l (\chi^1)_{\al},~~ -\bar{(\chi^1)}^{\hs{0.2mm}\dot{\al }}\r^T~.
\la{combin}
\eeq
As we see, $\psi_{+}$ and ${\rm i}\ga_5\psi_{-}$ are 4D Majorana spinors.
This will be used for constructing superfield's fermionic components 
by the 5D spinors. 

The superspace coordinates' fermionic component is the Majorana spinor
$\Theta =\l \te_{\al},~~ \bar{\te}^{\hs{0.2mm}\dot{\al }}\r^T~$. 
The superfield $H$ of eq.\eqref{hyperdec} reads
\beq
H=\phi_H+\ov{\Theta }_R(\psi_H)_R+\ov{\Theta }_R\Theta_R F~,
\la{H4comp}
\eeq
where $\ov{\Theta }_R(\Psi_H)_R={\Theta }_R^TC_4(\Psi_H)_R=
(\bar \te \bar \psi_H)$ and therefore $H$ is the chiral superfield with right 
chirality in two component notation.  From it we can build the superfield 
with left chirality $H=\phi_H^*-\te \psi_H-\te^2F^*$. With
$(\phi_H, \psi_H)=({\cal A}_2^{2\hat{\al }},~ 
(\psi_H)_R=-2{\rm i}\zeta^{2\hat{\al }}_R)$, we will have in two 
component notation
\beq
H={\cal A}_2^{\hs{0.3mm}2\hat{\al }*}+2{\rm i}(\te \eta^1)-\te^2F^*~.
\la{tilH2com}
\eeq
Similarly, from $H^c$ we can build superfield  with  left 
chirality in two component notation:
\beq
H^c={\cal A}_2^{\hs{0.3mm}2\hat{\al }-1*}-2{\rm i}(\te \eta^2)-
\te^2F^{c*}~.
\la{tilHc2com}
\eeq

In the actions \eqref{eq:vecsuperlagr},\eqref{hypAct} 
the superfield ${\mathbb W}_y$ is used which in
4-component notations is given by
$({\mathbb W}_y)_{4-comp}=e^5_y+2\ka \ov{\Theta }\ga^4\psi_{y-}+\cdots $.
In two component spinor notations, it is given by
\beq
{\mathbb W}_y=e^5_y\l 1+2\ka (\te \chi^1)+
2\ka (\bar \te \bar{\chi^{\hs{0.3mm}1}})\r 
+\cdots
\la{W2com}
\eeq
Here we also present the vector and (its partner) chiral superfields
$V, \Si $ in two component notation
$$
V=-(\te \si^{\bar \mu }\bar \te)W_{\bar \mu}+
2{\rm i}\te^2 (\bar \te \bar{\om^2})-2{\rm i}\bar \te^2 (\te \om^2)+
\fr{1}{2}\te^2\bar \te^2\l 2Y^3-\hat{D}_5M-
{\rm i}\pl_{\bar \mu }W^{\bar \mu }\r ~,
$$
\beq
\Si =\fr{1}{2}(e^5_yM+{\rm i}{W}_y)+
2e^5_y\te^{\al }\l {\rm i}\om^1_{\al }+\ka M\chi^1_{\al }\r
-\te^2F_{\phi }^*~.
\la{VSi2com}
\eeq
{}From $V$ the chiral superfield strength
$W_{\al }=-\fr{1}{4}\ov{D}^2D_{\al }V$  can be constructed. 
Also, the  superfield
${\cal V}_5$ in two component notations should be constructed trough
the combination
${\cal V}_5=(\Si +\Si^{\dagger }-\pl_yV)/{\mathbb W}_y$ with superfields
$\Si ,V, {\mathbb W}_y$ taken in two component spinor notation.

In the superspace actions eqs.\eqref{eq:vecsuperlagr} and \eqref{hypAct}, all 
superfields are assumed to be in the two component notations given in this 
appendix.

\renewcommand{\theequation}{D.\arabic{equation}}\setcounter{equation}{0}
\section{Warped backgrounds}\label{sec:warped}

In this appendix we show in detail how to deal with a warped background. As pointed out in section \ref{sec:gaugesugra}, the most straightforward approach is to use \emph{Weyl} transformations to reach a flat metric background. To be precise let us consider the following metric:
\be
            ds^2=a^2(y)dx_{\mu}dx^{\mu}-(e^5_y)^2dy^2,
\ee
where $e^5_y$ is a constant. If we perform a Weyl transformation 
\be
            e^a_{\mu}\to (e^a_{\mu})'=e^{-\sigma(y)}e^a_{\mu},
\ee
with $\sigma=\ln a$, we obtain a flat $4D$ metric. Under this transformation, the chiral multiplets arising from the hypermultiplets transform as
\be
             H\to H'=(\mcA',~\lambda',F')=e^{3\sigma/2}{\tilde H}, \quad\textup{with } {\tilde H}=(\mcA,~e^{\sigma/2}\lambda,~e^{\sigma}F),
\ee
while for the chiral multiplet which comes from the gauge sector, $\Sigma$, we have
\be
             \Sigma\to \Sigma'=(\phi,~e^{\sigma/2}\chi,~e^{\sigma}F_{\phi}).
\ee
In a similar way we obtain:
\be
             V\to V'=e^{\sigma}{\tilde V},\quad\textup{with } {\tilde V}=(W_{\mu},~e^{\sigma/2}2\Omega_+,~e^{\sigma}D),
\ee
\be
             \mcW_{\alpha}\to\mcW_{\alpha}'=e^{3\sigma/2}{\tilde \mcW_{\alpha}},\quad\textup{with } {\tilde \mcW_{\alpha}}=(-e^{\sigma/2}i2\Omega_{+\alpha},e^{\sigma}\left\{-i({\Fbarhat})_{\alpha}^{\beta} + \delta_{\alpha}^{\beta}(2Y^3-{\hat\mcD}_5 M)\right\},\cdots),
\ee
and 
\be
             {\mathbb W}_y\to{\mathbb W}_y'=e^{-\sigma}{\tilde{\mathbb W}_y},\quad\textup{with } {\tilde{\mathbb W}_y}=(~e^5_y,-e^{\sigma/2}2\ka\psi_{y-},-e^{\sigma}2\ka V_y^2,~e^{\sigma}2\ka V_y^1,\cdots).
\ee
We can use the Lagrangians presented in the bulk of the paper replacing the \emph{unprimed} fields by the \emph{primed} ones. This procedure leads to
\be
           \mcL_V=\frac{1}{4}\int d^2\theta\left( e^{3\sigma}P_{IJ}(\Sigma'){\tilde \mcW^{\alpha}}{\tilde \mcW_{\alpha}}+\cdots  \right)+\textup{h.c.}+\int d^4\theta\, e^{2\sigma}{\tilde{\mathbb W}_y}2P({\tilde \mcV}_5),
\ee
where
\be
           {\tilde \mcV}_5=\frac{\Sigma'+{\Sigma'}^+ -\partial_y V'}{\tilde{\mathbb W}_y}+\cdots  
\ee
Let us now consider the hypermultiplet Lagrangian. Without loss of generality, we will focus on case {\bf 1}. We get
\be\begin{split}
            \mcL_H= &\pm \int d^4\theta\,e^{2\sigma} {\tilde {\mathbb W}}_y\,2\left(   {\tilde H}^{\dagger}e^{-2gV'}{\tilde H}+{\tilde H}^{c\dagger}e^{2gV'}{\tilde H}^c   \right)\\
                    & \qquad-\int d^2\theta\, e^{3\sigma}\left( {\tilde H}^c\partial_y {\tilde H}-{\tilde H}\partial_y {\tilde H}^c-2g\Sigma' {\tilde H}^c {\tilde H} \right) +\textup{h.c.}
\end{split}\ee
Since the hypermultiplet can be either physical or a compensator, we introduced the pre-factor $\pm$.  

In the following we will show that the action of $\partial_y$ on the several powers of $e^{\sigma(y)}$ present in the Lagrangians doesn't lead to any terms absent in \cite{kugo01a}. To be more precise, the terms dependent on $\partial_y\sigma$ turn out to cancel out up to the terms arising from the 5D Ricci scalar. This will be shown for (part of) the bosonic part. Let us start with hypermultiplet Lagrangian, which in terms of components reads
\be\begin{split}\label{eq:onshellwarphyper}
        \mcL_H= & e^{4\sigma}e^5_y\bigg[\pm 2\left(|F|^2+|F^c|^2+ 2gD(|\mcA^c|^2-|\mcA|^2)\right)+\bigg(2F\left(\partial_y\mcA^c+\tfrac{3}{2}\mcA^c \partial_y\sigma+g e^5_y(M-iA_5)\mcA^c\right)\\
             & \hspace{120pt}+\left(F\to F^c,~\mcA\to -\mcA^c,~g\to -g \right)+\textup{h.c.}\bigg)\bigg].  
\end{split}\ee
After integrating out the $F$'s this can be rewriten as
\be\begin{split}\label{eq:Cwarpedhyper}
         \mcL_H=&\pm e^{4\sigma}e^5_y\bigg[-2\Big(|\nabla_5\mcA|^2+|\nabla_5\mcA^c|^2+g^2 M^2\left(|\mcA|^2+|\mcA^c|^2\right)-g(D+\partial_5 M)\left(|\mcA^c|^2-|\mcA|^2\right) \Big)\\
                & \quad -3(\partial_5\sigma)\partial_5\left(|\mcA|^2+|\mcA^c|^2 \right)+2gM(\partial_5\sigma)\left(|\mcA^c|^2-|\mcA|^2\right)-\frac{9}{2}(\partial_5\sigma)^2\left(|\mcA|^2+|\mcA^c|^2 \right)\bigg].
\end{split}\ee
Note that we assembled the $\partial_5\sigma$ terms in the second line of this equation. These terms are of two different types. There are two terms which are universal, i.e., they are equal for all hypermultiplets. After performing the sum over all hypermultiplets (including compensators) we get the following:
\be
      \mcL_H\supset -e^{4\sigma}e^5_y\left(\frac{3}{2}(\partial_5\sigma)\partial_5+\frac{9}{4}(\partial_5\sigma)^2\right)\mcA^2,          
\ee
where $\mcA^2\equiv 2\sum (-1)^d(|\mcA|^2+|\mcA^c|^2)$, with $d=0,1$, for physical and compensator hypermultiplets, respectively. The third term is non-universal, being present only for hypermultiplets charged under the symmetry gauged by $W_{\mu}$. This term must be canceled by a contribution coming from the vector sector. 

Let us consider now the vector sector. The Lagrangian contains
\be\label{eq:Cwarpedvector}
            \mcL_V\supset -e^{4\sigma}e^5_y \left[ \frac{1}{4}\mcN_{IJ}(M) D^I D^J+(\partial_5\sigma)\mcN_I(M) D^I+\frac{1}{2}\mcN_{IJ}(M) D^I \partial_5 M^J \right].
\ee 
Denote by $M^{I=i}$ the scalar of the vector multiplet ${\mathbb V}^{I=i}$ which couples to one of the hypermultiplets. The total Lagrangian involving the auxiliary fields $D^I$ is obtained from eqs.\eqref{eq:Cwarpedvector} and \eqref{eq:Cwarpedhyper}
\be
                 \mcL_D=-e^{4\sigma}e^5_y\left[ \frac{1}{4}\mcN_{IJ}(M) D^I D^J+\frac{1}{2}D^I R_I \right],
\ee
with
\be
                 R_I=\mcN_{IJ}\partial_5 M^J+2(\partial_5\sigma)\mcN_I-4\delta^i_I g(|\mcA^c|^2-|\mcA|^2).
\ee
After integrating out the $D^I$'s one gets a term $\propto (\partial_5\sigma)M^i(|\mcA^c|^2-|\mcA|^2)$ that exactly cancels the non-universal one in the second line of eq.\eqref{eq:Cwarpedhyper}. The total Lagrangian, which combines the vector and the hyper parts, is thus 
\be\begin{split}
                  \mcL_{H+V}\supset & e^{4\sigma}e^5_y\bigg[\frac{1}{4}\mcN_{IJ}\partial_5M^I\partial_5M^J -2\Big(|\nabla_5\mcA|^2+|\nabla_5\mcA^c|^2+g^2 {(M^i)}^2\left(|\mcA|^2+|\mcA^c|^2\right)\Big)\\
                              &\hspace{72pt} +4g^2 \mcN^{ii}\left( |\mcA^c|^2-|\mcA|^2\right)^2 -\left(\mcN-\frac{3}{2}\mcA^2\right)\left(\frac{5}{2}(\partial_5\sigma)^2+\partial_5^2\sigma \right) \bigg].
\end{split}\ee  
We can use now the fact that on-shell $\mcA^2=-2\mcN$. The last term in the equation becomes
\be
              e^{4\sigma}e^5_y\left(-2\mcN\left(5(\partial_5\sigma)^2+2\partial_5^2\sigma\right)\right).
\ee
This term arises from the 5D Ricci scalar:
\be
               {\mathcal R}\supset 4\left(5(\partial_5\sigma)^2+2\partial_5^2\sigma\right).
\ee
%Notice that in our derivation we have integrated out the auxiliary fields, but one can also show that the off-shell Lagrangians, eqs.\eqref{eq:onshellwarphyper} and \eqref{eq:Cwarpedvector}, coincide with the off-shell expressions of FKO if we take for $D^I$ and $F$,$F^c$ . 

\vspace{48pt}

\end{document}